\documentclass[10pt, conf, twocolumn]{IEEEtran}
\usepackage{color}
\usepackage[normalem]{ulem}
\useunder{\uline}{\ul}{}
\usepackage{times}
\usepackage{epsfig}
\usepackage{graphicx}
\usepackage{epstopdf}
\usepackage{amsmath}
\usepackage{cleveref}
\usepackage{amssymb}
\usepackage{amsxtra}
\usepackage{multirow}
\usepackage{mathtools}
\usepackage{caption}

\usepackage{url}
\usepackage{amsthm}
\usepackage{float}
\usepackage{subfig}
\usepackage{algorithmicx}
\usepackage{algorithm}
\usepackage[noend]{algpseudocode}

\algnewcommand{\Inputs}[1]{%
  \State \textbf{Inputs:}
  \Statex \hspace*{\algorithmicindent}\parbox[t]{.8\linewidth}{\raggedright #1}
}
\algnewcommand{\Initialize}[1]{%
  \State \textbf{Initialize:}
  \Statex \hspace*{\algorithmicindent}\parbox[t]{.8\linewidth}{\raggedright #1}
}

\newtheorem{theorem}{Theorem}
\newtheorem{lemma}{Lemma}

\newtheorem{corollary}{Corollary}

\newtheorem*{proof*}{Proof}

\makeatletter
\def\BState{\State\hskip-\ALG@thistlm}
\makeatother

\begin{document}

\title{%D2D Botnet in Wireless IoT Networks: Propagation Modeling and Optimziation under Advanced Persistent Threats
\vspace{0.0in}\hspace{-0.1in}Modeling, Analysis, and Mitigation of Dynamic Botnet Formation in Wireless IoT Networks}

\author{ \IEEEauthorblockN{\large Muhammad Junaid Farooq, \textit{Student Member, IEEE} \\  and Quanyan Zhu, \textit{Member, IEEE} }
%\\ \IEEEauthorblockA{Department of Electrical \& Computer Engineering, Tandon School of Engineering, \\New York University, Brooklyn, NY 11201, USA,} Emails: \{mjf514, qz494\}@nyu.edu. \vspace{-0.25in}
%{\thanks {\vspace{-0.2cm}\hrule \vspace{0.2cm} \indent This work was made possible by.}}

\thanks{\vspace{-0.0in}\hrule \vspace{0.2cm}
%A part of this work has been submitted to 56th Annual Allerton Conference on Communication, Control, and Computing~\cite{junaid_allerton}.

Muhammad Junaid Farooq and Quanyan Zhu are with the Department of Electrical \& Computer Engineering, Tandon School of Engineering, New York University, Brooklyn, NY, USA, E-mails: \{mjf514, qz494\}@nyu.edu. \vspace{-0.0in}}
}

\maketitle

\begin{abstract}
The Internet of Things (IoT) relies heavily on wireless communication devices that are able to discover and interact with other wireless devices in their vicinity. The communication flexibility coupled with software vulnerabilities in devices, due to low cost and short time-to-market, exposes them to a high risk of malware infiltration. Malware may infect a large number of network devices using device-to-device (D2D) communication resulting in the formation of a botnet, i.e., a network of infected devices controlled by a common malware. A botmaster may exploit it to launch a network-wide attack sabotaging infrastructure and facilities, or for malicious purposes such as collecting ransom. In this paper, we propose an analytical model to study the D2D propagation of malware in wireless IoT networks. Leveraging tools from dynamic population processes and point process theory, we capture malware infiltration and coordination process over a network topology. The analysis of mean-field equilibrium in the population is used to construct and solve an optimization problem for the network defender to prevent botnet formation by patching devices while causing minimum overhead to network operation. The developed analytical model serves as a basis for assisting the planning, design, and defense of such networks from a defender's standpoint.
\end{abstract}

\IEEEpeerreviewmaketitle
\vspace{-0.0in}
\begin{IEEEkeywords}
Botnet, Internet of Things, device-to-device communication, population processes, distributed denial of service.
\end{IEEEkeywords}

\vspace{-0.0in}
\section{Introduction}

%\textcolor{red}{
%\begin{itemize}
%\item System model figure
%%\item Plots of approximation accuracy of Equilibrium characterization
%%\item plots of minimum function to show accuracy
%%\item CoV approach to solve the infinite dimensional problem
%%\item Add proofs of convexity as lemma and derive conditions for it
%%\item Section on Game Theoretic optimization
%%\item Describe algorithm for dual problem solution
%%\item Mention attacker capabilities
%%\item Duality gap and algorithmic convergence
%\end{itemize}
%}

The Internet of things (IoT) comprises of a network of sensors and actuators, which are embedded computers, communicating with each other and to the Internet. Often, the endpoint devices rely on a plethora of wireless communication technologies and protocols such as WiFi, Bluetooth, Zigbee, etc.,~\cite{iot_ran}. Although most devices in an IoT network are directly connected, via access points, to the Internet; there is an inherent flexibility in devices to connect to other wireless devices in their communication range in order to leverage their capabilities resulting in powerful functionalities. Furthermore, some commercially available devices have extremely versatile processing and communication capabilities, e.g., the Amazon Echo~\cite{echo}, Google Home~\cite{g_home}, etc., which enables them to execute custom programs and processes.
% a significant portion of the network uses peer-to-peer (P2P) wireless communication based on a variety of protocols including WiFi, Bluetooth, Zigbee, etc. such as between hubs and routers.

IoT devices are manufactured by different vendors without strong regulations on embedding cyber security features in the software. To reduce cost and time-to-market, security issues may be overlooked by device manufacturers~\cite{flaws}. In addition to inherent software vulnerabilities, several other factors increase the risk of cyber attacks on these devices. One of the risks is the use of stock passwords to access the control panel of these devices. Moreover, most IoT devices are left to operate on consumer premises without regular maintenance. It exposes them to the risk of being infected and controlled by malicious software processes, referred to as \emph{malware}~\cite{cyber_security}. It is also possible that consumers might willingly accept to install certain processes or applications on their devices in return for financial incentives, completely unaware of the fact that they might be used to launch a distributed denial of service (DDoS) attack~\cite{ddos_in_iot} on the network at a later stage.

%Furthermore, user unawareness of security issues is also an alarming concern. The propagation of botnets. This is an alarming concern to the rapidly growing IoT industry to avoid distributed denial of service (DDoS) attacks~\cite{ddos_in_iot}.

  %On the other hand, some applications may be recruiting botnets in disguise. The future of IoT world lies in data monetization and  it is possible that applications pretending to pay devices for data collection may actually act as bots recruited by the botmaster.

Botnets have become a significant threat to computer and communication networks in the last decade~\cite{botnet}. A botnet is a network of devices infected by malicious software and controlled by an external operator referred to as the \emph{botmaster}~\cite{botnet_survey_ppr}. Often, the malware infiltrates the network stealthily over time in a self-replicating manner before being instructed by the botmaster to trigger an attack. The objective of the botnet is to cause disruption in service provisioning leading to loss of operation and sometimes with the intent of obtaining ransom~\cite{ransom}. The most famous  botnet attack in recent history has been the \emph{Mirai} in 2016~\cite{mmirai}. Recently, researchers have identified variants of the Mirai botnet referred to as the \emph{IoTroop} or \emph{Reaper} that is aimed at using IoT devices to launch DDoS attacks~\cite{variant}. It is a powerful botnet that comprises of compromised domestic wireless routers, TVs, DVRs, and surveillance cameras exploiting vulnerabilities in devices from major manufacturers.

In the case of wireless IoT networks, the malware may spread from one device to another among devices that are in close geographical proximity~\cite{wifi_botnets}. Due to the absence of centralized connectivity, the botmaster is compelled to use the same D2D links to issue control commands for coordinating an attack. Seed viruses may be planted into the networks using malicious or infected IoT devices or even using UAVs~\cite{drone_botmaster}. Moreover, the botmster may change the malware code dynamically and may issue control commands to launch a wireless denial of service (WDoS) attack~\cite{wdos}. It is different from traditional DDoS attacks as services do not have to be taken off the Internet. Instead, the goal is to exploit MAC vulnerabilities in wireless devices to generate superfluous traffic that sabotages legitimate operation~\cite{mac_whitepaper}. The D2D nature of the wireless communication network makes it harder to launch a coordinated DDoS. However, at the same time, it is also hard to defend against it as a network of devices contributes to the attack and there is no single source. Therefore, the best strategy for a network defender is to prevent the dynamic development of a large scale botnet and limit its ability to launch a DDoS.

Several dynamic processes might be burgeoning in the network at the same time. Malware in an infected device might be attempting to replicate itself in nearby devices. Furthermore, the infected devices also share control commands with other infected devices to agree on an attack point. On the other hand, the network defense mechanisms are also in place which periodically patch\footnote{Throughout this paper, the term `patch' refers to attempts made by the defender to bring the device to an un-compromised state, e.g., via power cycling, firmware upgrades, etc.} the devices. The patching frequency of devices needs to be carefully selected as it negatively affects the regular device operation. Particularly, if a device acts as a hub, i.e., connecting multiple devices together, the impact of downtime will be much more severe. In order to make such optimal patching frequency decisions, we need a theoretical model that can accurately capture the connectivity characteristics of the network and incorporate the continuing dynamic processes. 
%The defender needs to patch the devices successfully at different rates based on the connectivity of devices to prevent the attack.

\textcolor{black}{While the modeling and analysis of traditional Internet based botnets is also important due to its huge monetary and non-monetary impact, there have been some efforts to prevent and control them. However, the botnets in wireless IoT systems need special attention due to the current lack of awareness and the increased security vulnerability of IoT devices}. Despite the impending security threat to a massive number of unprotected IoT devices and systems, there is a severe dearth of systematic methodologies for understanding such systems from a security standpoint. 
%Although the problem of developing optimal strategies to prevent botnet formation in wireless IoT networks is highly important, however, there are still a shortfall of concrete theoretical models available to study such scenarios and develop preemptive and resilient mechanisms to counter the negative outcomes. 
This necessitates the development of exclusive models for such wireless IoT networks which can capture the spatial distribution of the devices and the dynamic processes of malware infiltration, control command propagation, and device patching by the defender. In this paper, we develop the theoretical underpinnings that allow the modeling and analysis of dynamic botnet formation in wireless IoT networks. A summary of the main contributions is provided below:
\begin{enumerate}
\item We propose a novel analytical model, inspired from the dynamics of population processes, to capture the dynamic formation of botnets in wireless IoT systems using D2D communication.
\item We analyze the degree based mean field equilibrium populations of malware-free devices and control command aware devices in the network and develop approximate tractable expressions for them.
\item We formulate an optimization problem from a network defender's standpoint, to control the formation of a botnet via patching while causing minimum disruption to regular operation of the IoT network, which turns out to be non-convex.
\item We prove that the formulated non-convex optimization problem has zero duality gap and consequently solve it using a dual decomposition based algorithm to obtain the optimal patching policy and study its behavior in response to varying network parameters.
%\item We propose an optimization framework that aims to determine the patching rate required for each class of device in the network to control the formation of a botnet while causing minimum disruption to regular operation of the IoT network.
%\item We develop a degree based mean field epidemic spreading model to analyze the propagation of malware from one device to another in its transmission range.
%\item We consider the development of distributed botmasters that control the activity of the bots while maintaining coordination among themselves for developing consensus on the attack strategy and timing.
%\item We use stochastic geometry to capture the probability of successful message transmission between devices due to channel impairments and interference.
%\item A two player game has been formulated between the attacker and defender to model the interaction between the botnet and network defender. Optimal attack and defense strategies in terms of infiltration and recovery rates have been obtained based on the best response of each player.
\end{enumerate}

The rest of the paper is orgainzed as follow: Section~\ref{Sec:Related_Work} provides a review of existing literature, Section~\ref{Sec:Syst_Model} provides a description of the system model including the network setup and threat model used. Section~\ref{Sec:Method} provides a detailed description on the modeling of malware \& information evolution in the network, state space representation \& dynamics, and equilibrium analysis. It also provides a formulation of the network defense problem and its solution methodology. Section~\ref{Sec:Results} provides results of numerical experiments and the corresponding analysis. Finally Section~\ref{Sec:Conclusion} concludes the paper with potential future research directions.

\vspace{-0.0in}
\section{Related Work}\label{Sec:Related_Work}
\textcolor{black}{In recent years, significant efforts have been invested in research on Botnets and their characteristics~\cite{soa3}. Most studies are focused on Internet botnets~\cite{tifs_2_internet} or, more specifically, on IP based networks~\cite{measurement}. However, the botnet phenomenon has been sparingly investigated in wireless networks. Furthermore, the existing studies are are either based on simulations~\cite{simulation,soa4} or use abstract theoretical models that do not capture the dynamics of malware propagation or the network geometry into account~\cite{iot_epidemic}.
%spread of malware in traditional wireless networks such as. 
In general, there is a lack of analytical modeling and analysis to support the frameworks developed developed particularly for malware spreading that may lead to a coordinated attack as in a botnet.
}
%lack of analytical modeling and analysis to support the results
%Although mobility is important, however, a large proportion of the IoT networks will be static deployed wireless networks such as wireless routers, TVs, DVRs, IP cameras. Therefore, the fundamental understanding of such behavior is highly important.

%Most of these works are in centralized wireless systems such as in cellular networks~\cite{cellular_botnets}. Some studies have been done is adhoc wireless setting. For instance, the propagation of viruses in wireless networks using the bluetooth interface is studied in~\cite{bluetooth_propagation}.

 %Although, the evolution of botnets in mobile adhoc networks due to the mobility of malware infected devices has been investigated in~\cite{tmc_botnets}, however, it uses a bond percolation approach which does not capture the dynamics over a topology of wireless devices.

\textcolor{black}{The most related research to our proposed work is presented in~\cite{tmc_botnets,soa1,soa2}. \cite{soa1} uses game theory and epidemiology to study security risks in D2D offloading of computational tasks between devices, \cite{tmc_botnets} investigates mobile botnets spreading infection in a D2D fashion on the go, and~\cite{soa2} considers the case when multiple bots are trying to attack a single server. While these are trying to mitigate the risks of a large scale DDoS by a botnet, they do not not account for the dynamics of the malware propagation or the network geometry aspects that are important in wireless IoT networks. On the other hand, a framework for preventing malware propagation in wireless sensor networks has been proposed in~\cite{tifs_differential_game} that captures the network features, however, it does not take into account the stealthy propagation behavior of a botnet which requires information dissemination and coordination to launch an attack. To overcome the challenge of understanding the propagation of information in wireless networks, we have proposed a framework in our previous works~\cite{twc_junaid, wiopt_junaid}, leveraging concepts from mathematical epidemiology~\cite{book_epidemiology} and point process theory. However, traditional epidemiological models such as~\cite{virus_spread} and~\cite{rumour_dynamics} are not sufficient to analyze the botnet formation in wireless networks due to the interplay between malware infection, control commands propagation, and device patching.\newline
\indent In this paper, we develop novel methodologies to overcome the unique challenges of modeling and analyzing the crucial interplay between malware infection, control commands propagation, and device patching in wireless IoT networks. We leverage ideas from the theories of dynamic population processes~\cite{population_processes} and point processes to setup a mean field dynamical system that captures the evolution of malware infected devices and control command aware devices over time.
In general, obtaining tractable characterizations of the equilibrium state in such population processes is theoretically involved due to the self-consistent nature of the equations involved and the complex connectivity profile of the network. In this paper, we propose a variation of the mean field population process model based on a customized state space that allows us to analyze the formation of botnets in wireless IoT networks and helps in making decisions to control its impact.
%Although it captures some aspects of the mobility of network nodes, however, it uses a bond percolation approach which does 
%which are crucial features for botnet formation in wireless IoT networks.
}

%It is important to highlight that in a wireless IoT botnet, both the malware and control signals of activation from botmasters use the same wireless interface to interact with the devices. The initial infiltration requires the exploitation of vulnerabilities in network devices. However, once the devices are compromised, the communication between them is only influenced by the wireless communication channel. Therefore, it is important to study the coupled interaction of worm spreading as well as control information propagation over such wireless IoT networks.

\vspace{-0.0in}
\section{System Model} \label{Sec:Syst_Model}

\begin{table}[]
	\centering
	\caption{List of model parameters.}
	\setlength\tabcolsep{3pt}
	\label{parameters}
	\begin{tabular}{|c|l|}
		\hline
		Symbol & Description                                                             \\ \hline
		$\lambda$ & Density of deployed devices modeled according to a PPP                  \\ \hline
		$r$       & Communication range of devices                                          \\ \hline
		$\rho$    & Probability of successful transmission between devices \\ \hline
		$p$    & Proportion of devices vulnerable to malware infiltration \\ \hline
		$K$    & Degree (number of communication neighbors) of a typical device \\ \hline
		$\pi_k$    & Probability that a typical device has degree $K=k$. \\ \hline
		$k_{\max}$    & Maximum possible degree in the network \\ \hline		
		$\gamma_b$    & Malware spreading rate of bot \\ \hline
		$\gamma_c$    & Control commands propagation rate of bots \\ \hline	
		$\sigma_1$    & Average probability of being connected to a bot device \\ \hline
		$\sigma_2$    & Average probability of being connected to an informed bot \\ \hline	
		$\theta_{\tilde{B}}$    & Probability that a given link points to an un-compromised device \\ \hline
		$\theta_{BI}$    & Probability that a given link points to an informed bot device \\ \hline
		$\beta$    & Information refresh rate of bot devices \\ \hline
		$\mu_k$    & Patching rate of device with degree $K = k$ \\ \hline
		$\tau_{\tilde{B}}$    & Minimum proportion of un-compromised devices in the network \\ \hline
		$\tau_{BI}$    & Maximum proportion of informed bots in the network \\ \hline
	\end{tabular}
\vspace{-0.0in}
\end{table}

%The botmaster may have to invest more resources in identifying new software vulnerabilities allowing it to intrude more computers.
%change the malware program to incorporate new
%On the other hand, the network defender needs to make efforts in potentially scanning, updating device firmware, and/or re-booting the devices.
%send more updates or require password changes to the customers/ equipment users.

In this section, we provide a description of the network model used and the associated threat model. For the convenience of readers, the notations used throughout this paper are summarized in Table~\ref{parameters} along with a brief description.

\vspace{-0.0in}
\subsection{Network Model}
\begin{figure}[t]
	\centering
	\includegraphics[width=3.2in]{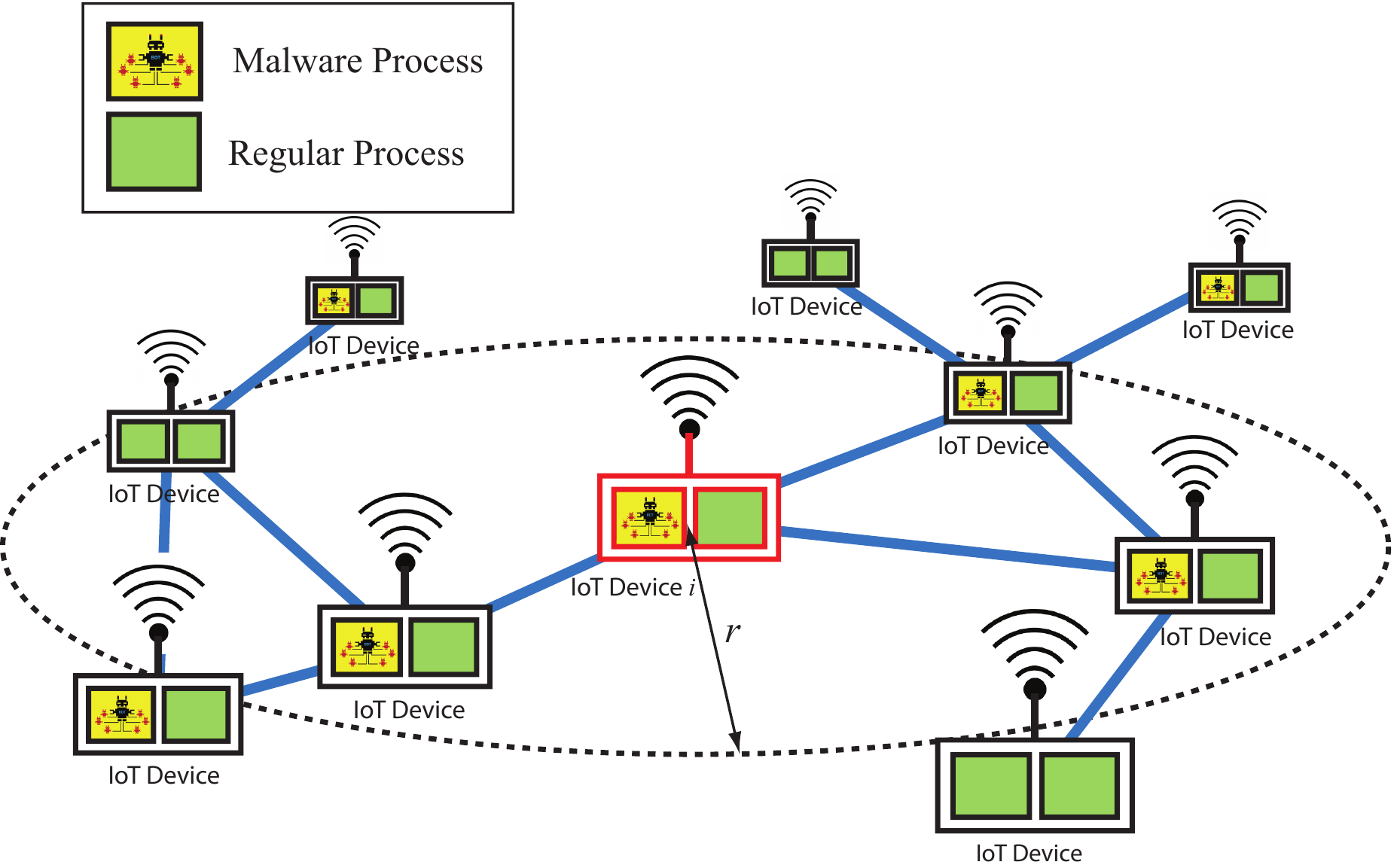}\\
	\caption{Network model: A typical IoT device, referred to as device $i$, is highlighted in red colour. Each IoT device executes a regular process (indicated by green boxes) and may or may not be running a malware process (indicated by the yellow boxes with a bot symbol if infected or gray box otherwise). Devices within the communication range (indicated by the dotted line for device $i$) of each other are assumed to be able to communicate with each other and the communication links are highlighted by blue lines between the devices.}\label{Fig:system_model}
\end{figure}

We consider a set of wireless IoT devices uniformly distributed in $\mathbb{R}^2$ according to a homogeneous Poisson Point Process (PPP)~\cite{sg} denoted by $\Phi = \{x_i\}_{i \geq 1}$ with intensity $\lambda \in \mathbb{N}$ devices/km$^2$, where $x_i \in \mathbb{R}^2$ represents the location coordinates of the $i^{th}$ device. Each device has computing capabilities for executing processes and has a wireless interface for communication with neighboring devices. The devices are assumed to have omni-directional transmissions with a communication range of $r$ m. A typical device located at $x_i$ is connected wirelessly with $K = |N_i|$ other devices, where $N_i = \{ j : \| x_i  - x_j \| \leq r , \ \forall j \neq i\}$ and $|.|$ denotes the cardinality operator. Since the devices in the network are distributed according to a PPP, the degree $K$ is a random variable with $\mathbb{P}[K = k] = \pi_k = \frac{e^{-\lambda \pi r^2}(\lambda \pi r^2)^k}{k!}$. Furthermore, the average degree of a typical device is $\mathbb{E}[K] = \lambda \pi r^2$. An illustration of the network setup along with the state at a particular time is provided in Fig.~\ref{Fig:system_model}. A realization of a random network is shown where each IoT device is shown to be equipped with a wireless interface and executing a regular process and a malware process (if infected). The device connectivity is represented by blue links between devices that are within a distance $r$ of each other. The malware and the control commands propagate over these wireless links from one devices to another. A simultaneously executing patching process restores the devices to an un-compromised state (illustrated by the gray boxes).

\begin{figure*}[t]
\centering
\subfloat[Location of WiFi hotspots in New York City.]{\includegraphics[width = 2.8in]{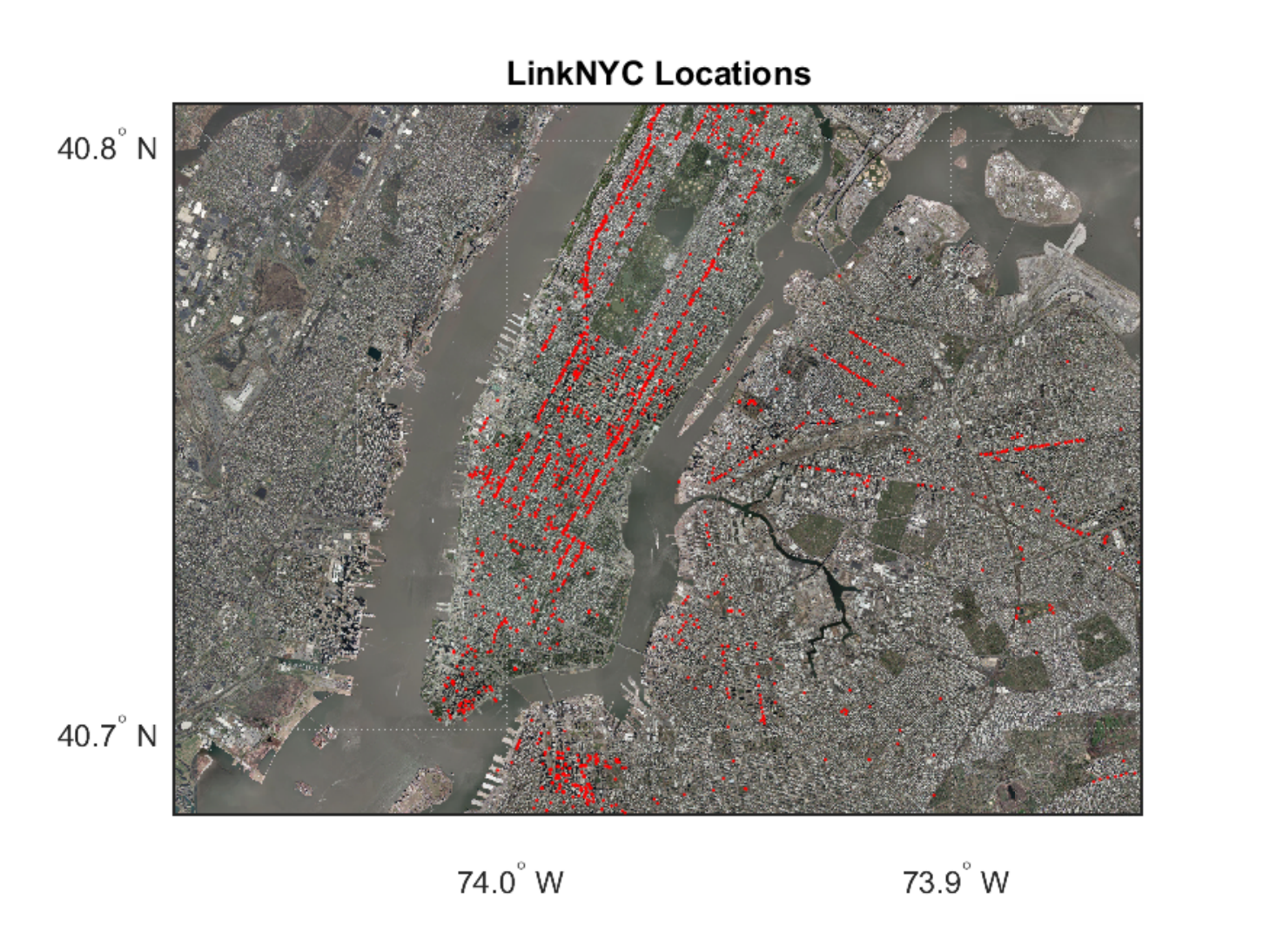} \label{Fig:map}} \ \
\subfloat[]{\includegraphics[width = 2.8in]{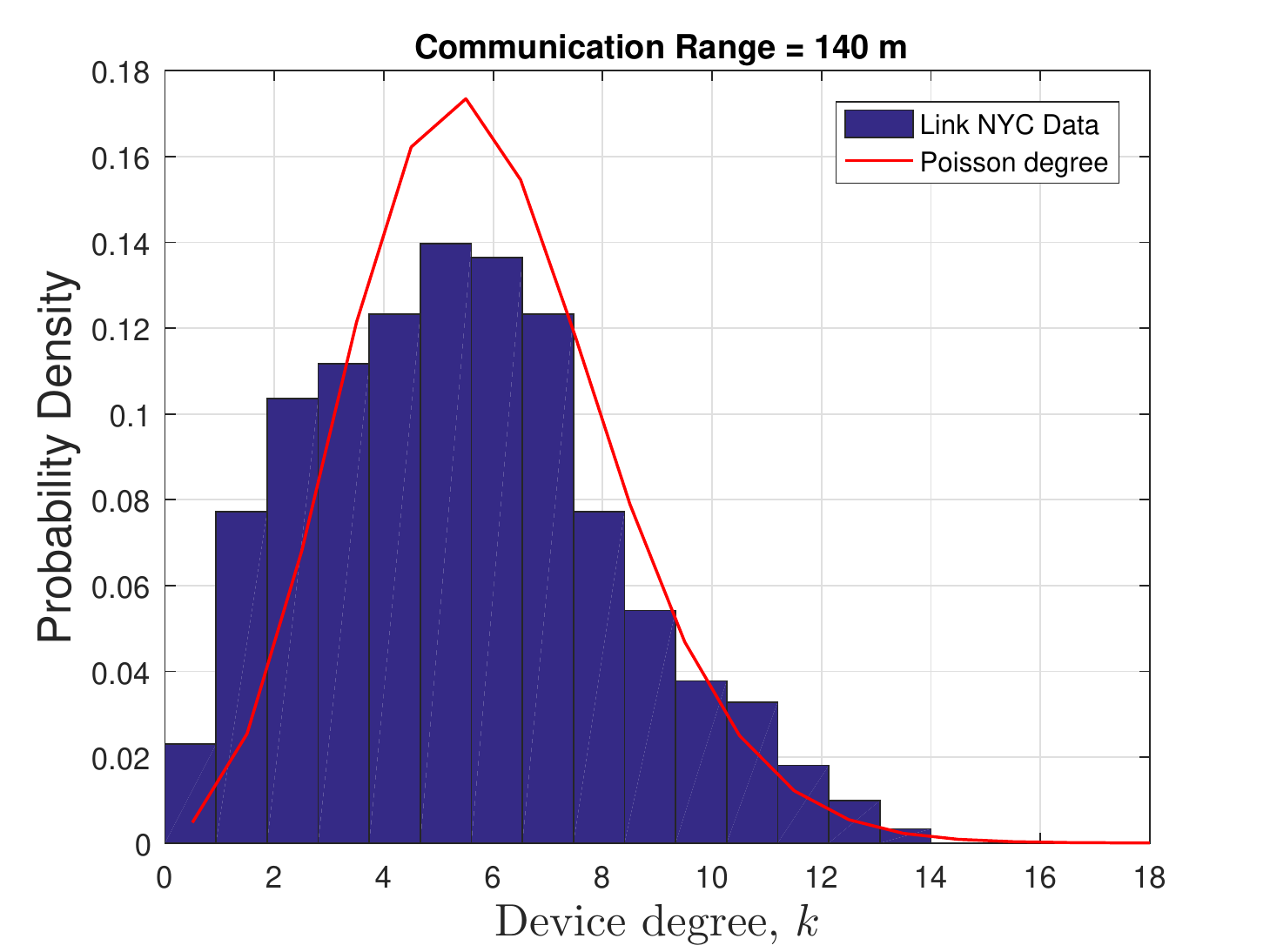} \label{Fig:poisson_fit}}
\caption{Analyzing potential connectivity of WiFi hotspots in NYC.}
\label{Fig:linknyc}
\end{figure*}

In order to demonstrate the practical applicability of the employed PPP network model and the associated degree profile of the devices, we use location data of WiFi access points in New York City (NYC), referred to as LinkNYC~\cite{nyc_hotspot_data}. A map of the locations of hotspots is provided in Fig.~\ref{Fig:map}. We use the locations data of 652 hotspots located in Midtown Manhattan and surrounding neighbourhoods. Assuming the wireless IoT devices are deployed at the locations of LinkNYC hotspots with a communication range of 140 m, the connectivity profile of a typical devices will almost be Poisson distributed\footnote{Note that the LinkNYC data has been used as an example to demonstrate the idea of wireless device reachability in large scale public/privately deployed IoT devices in the future.}. The empirical degree distribution along with the maximum likelihood estimated Poisson degree is shown in Fig.~\ref{Fig:linknyc}. Some distortion is observed due to the physical limitation on the hotspots to be confined to the Manhattan grid lines.

We assume that the network is uncoordinated and the devices communicate with each other using (ALOHA)~\cite{aloha} as the medium access control (MAC) protocol. In other words, the devices do not coordinate with each other in making transmission decisions\footnote{Note that the subsequently proposed framework is not restrictive to a particular MAC protocol. Other MAC protocols such as the carrier sense multiple access (CSMA) can also be used, however, the mean-field dynamics may not directly apply.}. A Significant amount of literature is available on capturing the effects of interference, characterizing the probability of transmission success, and evaluating transmission capacity in Poisson wireless ad hoc networks~\cite{aloha_SG}. In this paper, we introduce the probability of transmission success of a typical transmitting device as a parameter $\rho \in [0,1]$. Precise characterization can be obtained using tools from stochastic geometry~\cite{sg}, such as in~\cite{outage_aloha,kaynia_outage}, however it is not the main focus of this work.

\vspace{-0.0in}
\subsection{Threat Model}
We assume that a botmaster, i.e., the entity which has authored the malware and subsequently plans to launch an attack, possesses powerful capabilities to exploit loopholes in vulnerable wireless IoT devices to infiltrate them and install malicious software process on them. We assume that a proportion $p \in [0,1]$ of the network is vulnerable to being compromised or infiltrated by the malware if the malware has been successfully transmitted over the wireless interface\footnote{Vulnerability to be compromised can emanate from events such as using default passwords for access control, using an older version of the firmware etc.}. In other words, $1/p$ can be considered to be the average number of successful transmission attempts required to infiltrate a neighboring device.

The bots use a fraction of the communication resources of the host device to infiltrate nearby devices and to share control commands. The transmission rate of packets to break into other devices is referred to as \emph{malware spreading rate} and denoted by $\gamma_b \geq 0$ in units of packets per second. Similarly, the transmission rate of packets contributing towards the dissemination of control commands is referred to as \emph{control command propagation rate} and denoted by $\gamma_c \geq 0$. Note that the sum of $\gamma_b$ and $\gamma_c$ must be sufficiently small in order to maintain stealthy operation of the botnet.

In summary, the botnet threat in the wireless IoT networks is two fold. Firstly, the malware may spread from one device to another in its proximity using the wireless interface. Secondly, the infected devices referred to as bots share control commands using the same wireless medium to coordinate and plan for launching a network-wide attack. However, as soon as a particular device is patched, the malicious process running on the device is terminated and it gets rid of both the malware as well as information about the control commands. After being patched, the device becomes vulnerable to infection again in the future\footnote{In practice, the device vulnerability for future infection may reduce after getting patched, however there always exists a certain minimum vulnerability level of the devices. Moreover, the botmaster may also update its strategies to render the devices vulnerable again.}.

\vspace{-0.0in}
\section{Methodology} \label{Sec:Method}

In this section, we provide a systematic approach to model the propagation of malware and formation of a botnet in wireless IoT networks. The proposed model is formally described using the dynamics of population processes and the analysis of equilibrium is presented. Finally, a network defense problem is formulated and a polynomial time algorithm is proposed to obtain the optimal device patching strategy mitigating the formation of a botnet and associated risk of network-wide attack.

\vspace{-0.0in}
\subsection{Modeling of Malware \& Information Evolution}

In a large scale wireless IoT network, a typical device may either be un-compromised or infiltrated by malware, thus referred to as a \emph{bot}. Furthermore, devices that are bots may or may not have received control commands. Those that have received control commands may have discarded them due being stale or outdated. Note that since the devices may go from one state to the other based on their communication interactions within their neighborhood, it is appropriate to categorize the devices according to their connectivity or degree\footnote{This implies that devices with similar connectivity profile will have similar behavior in terms of botnet fromation.}. This allows us to use the degree based mean field approach to study the spread of malware and their communication~\cite{epidemics}. The possible system states of the population of degree $k$ devices, i.e., devices that are capable of communicating with $k$ other devices, can then be classified as follows:
\begin{itemize}
	\item $\tilde{B}_k$ - the proportion of degree $k$ devices in the network that are un-compromised.
	\item $B\tilde{I}_k$ - the proportion of degree $k$ devices in the network that are bots but uninformed about control commands.
	\item $BI_k$ - the proportion of degree $k$ devices in the network that are bots and are also informed with control commands.
\end{itemize}

\vspace{-0.0in}
Once, the states are defined, we can study the transitions between each of these states. At any given time an un-compromised device may become a un-informed bot at a rate that it proportional to its degree $k$ and the average probability that it is connected to a bot device, denoted by $\sigma_1$. Similarly, an un-informed bot may become an informed bot at a rate that is proportional to its degree $k$ and the average probability that it is connected to an informed bot, denoted by $\sigma_2$. On the other hand, an informed bot may discard the control commands at a constant rate $\beta$ to return to an un-informed state to maintain recency of control information. Finally, if the bots are patched, they return to an un-compromised state. We use a degree based patching rate $\mu_k$ inspired from the non-uniform transmission model proposed in~\cite{nonuniform_transmission}. This completes all the transitions between the possible system states.

%\subsection{State Evolution \& Dynamics}

%\textcolor{red}{The parameter $\beta$ is used to model how aggressive the bots are towards exchanging the most most recent information in the network. In other words, it models how quickly the bots in informed sate will disregard the control messages. A higher rate indicates a quicker transition from informed to uninformed state.}

%\textcolor{red}{Key Assumption: We assume that the device vulnerability does not change after each recovery. We can use it as a special case to the general case when device vulnerabilities change every time it is patched.}

%\textcolor{red}{Degree dependent spreading rate has been investigated in using $\gamma(k) = \gamma_0 \frac{ k^{\alpha}}{k}$. }

%\textcolor{red}{Describe the states and their transitions. The patching behaviour etc.}
%At any given time, any degree $k$ device in the network may be in one of the following states: (i) $\tilde{B}$
%The states also depict the population evolution over time in the network.
%\textcolor{red}{The recovery is independent of the malware spread or the communication. The dependence in the communication between bots on the infiltration is captured in the state diagram.}

\vspace{-0.0in}
\subsection{State Space Representation \& Dynamics}
\begin{figure}
	\centering
	\includegraphics[width=3in]{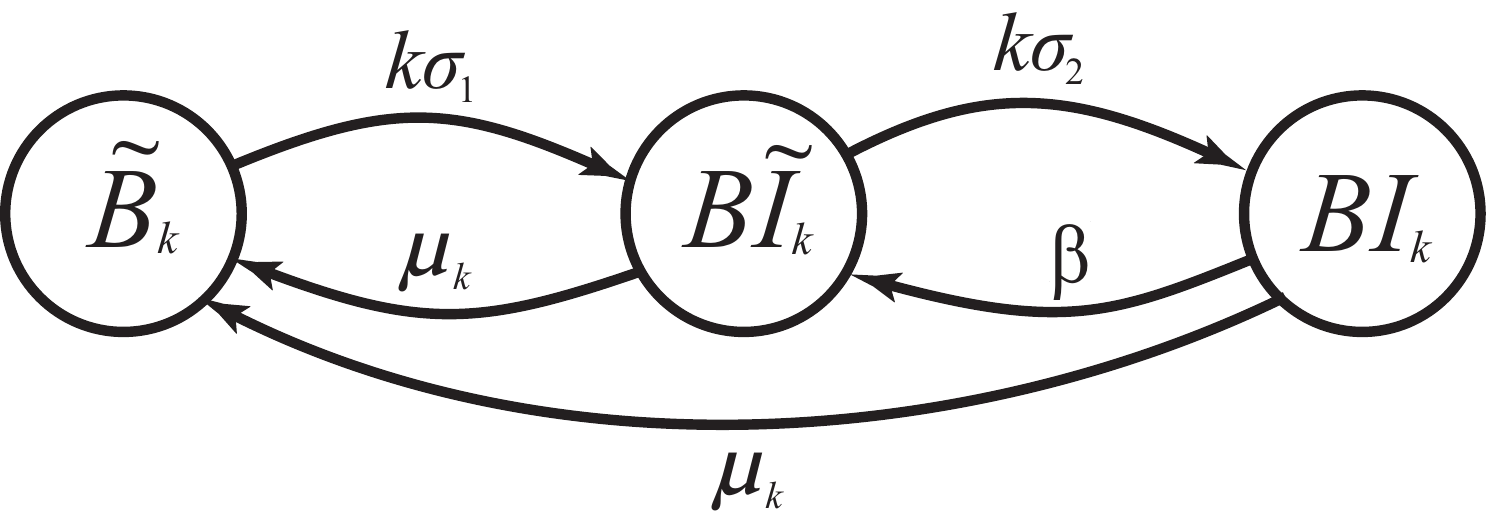}\\
	\caption{State evolution diagram for a typical device. Un-compromised devices of degree $k$, represented by $(\tilde{B}_k)$ may become infected with malware to become un-informed bot devices ($B\tilde{I}_k$), which can further become informed bots ($BI_k$). The informed devices discard information at a rate $\beta$ to again become un-informed. A patching process brings both un-informed and informed bots to an un-compromised state.\vspace{-0.0in}}\label{Fig:state_diagram}
\end{figure}

In this subsection, we formally express the dynamics of the system using the developed state space.
The state space representation and associated transitions described in the previous subsection are illustrated by the state diagram shown in Fig.~\ref{Fig:state_diagram}.
Using the figure and leveraging concepts from the theory of population processes~\cite{population_processes}, the state evolution can be mathematically described by the following dynamical system of equations:
\begin{align}
& \hspace{-0.1in}\frac{d\tilde{B}_k(t)}{dt} = \mu_k (B\tilde{I}_k(t) + BI_k(t)) - k \sigma_1 \tilde{B}_k(t), \notag \\
& \hspace{-0.1in} \hspace{1.1cm} = \mu_k (1 - \tilde{B}_k(t)) - k \sigma_1 \tilde{B}_k(t), \label{Eq:Orig1} \\
& \hspace{-0.1in}\frac{dB\tilde{I}_k(t)}{dt} = -(\mu_k \hspace{-0.02in} + k \sigma_2) B\tilde{I}_k(t)\hspace{-0.02in} + k \sigma_1 \tilde{B}_k(t) + \beta BI_k(t), \label{Eq:Orig2} \\
& \hspace{-0.1in}\frac{dBI_k(t)}{dt} = - (\mu_k + \beta) BI_k(t) + k \sigma_2 B\tilde{I}_k(t). \label{Eq:Orig3}
\end{align}
Note that~\eqref{Eq:Orig1} captures the birth and death processes of un-compromised devices. In other words, it implies that at time $t$, the population proportion of un-compromised degree $k$ devices is increasing with a rate that is proportional to the patching rate and the population proportion of bot devices. However, at the same time, it is also decreasing at a rate that is proportional to the degree $k$ and the expected rate of interacting with a bot device. Similarly, we can interpret the remaining dynamical equations for un-informed bot and informed bot populations. Since, the states represent the population proportions, we can use the closure relationship, i.e., $\tilde{B}_k(t)~+~B\tilde{I}_k(t)~+~BI_k(t)~=~1, \forall t \geq 0$, to reduce \cref{Eq:Orig1,Eq:Orig2,Eq:Orig3} to the following independent dynamical system of equations:
\begin{align}
&\hspace{-0.1in}\frac{d \tilde{B}_k(t)}{dt} = \mu_k - (\mu_k + k \sigma_1) \tilde{B}_k(t), \label{Eq:ind_dyn1}\\
&\hspace{-0.1in}\frac{d BI_k(t)}{dt} = k \sigma_2 - (\mu_k + \beta + k \sigma_2) BI_k(t) - k \sigma_2 \tilde{B}_k(t) \label{Eq:ind_dyn2}.
\end{align}
%$\sigma_1$ is the average rate of an uncompromised device to be infiltrated by malware and become a bot. $\sigma_2$ is the average rate of an uninformed bot to become informed with botmaster messages. These probabilities can be evaluated as follows:
Note that the average probability for a degree $k$ device to be connected to a bot device, $\sigma_1$ is directly proportional to the probability of transmission success, the vulnerability of the devices, the malware spreading rate, and the probability of being connected to a bot device. Similarly, the the average probability for a degree $k$ device to be connected to an informed bot, $\sigma_2$ is directly proportional to the probability of transmission success, the control command propagation rate, and the probability of being connected to an informed bot device. These can be, respectively, expressed as follows\footnote{The event of a device being vulnerable to malware infection and the successful reception of wireless signals are independent. Hence, the probabilities can be directly multiplied.}:
\begin{align}
&\sigma_1 = \rho \gamma_b p (1 - \theta_{\tilde{B}}),\\
&\sigma_2 = \rho \gamma_c \theta_{BI},
\end{align}
where $\theta_{\tilde{B}}$ is the probability that a particular link of a degree $k$ device points to an un-compromised device, and $\theta_{BI}$ is the probability that a particular link of a degree $k$ device points to an informed bot device. These probabilities can be evaluated as $\theta_{\tilde{B}} = \sum_{k^{\prime}} \mathbb{P}(k^{\prime}|k) \tilde{B}_{k^{\prime}}(t)$ and $\theta_{BI} = \sum_{k^{\prime}} \mathbb{P}(k^{\prime}|k) BI_k(t)$. However, for networks with uncorrelated degrees, these probabilities can be further expressed as follows:
\begin{align}
&\theta_{\tilde{B}} = \sum_{k^{\prime}} \frac{k^{\prime} P(k^{\prime})}{\mathbb{E}[K]} \tilde{B}_{k^{\prime}}(t), \label{Eq:theta_1}\\
&\theta_{BI} = \sum_{k^{\prime}} \frac{k^{\prime} P(k^{\prime})}{\mathbb{E}[K]} BI_{k^{\prime}}(t).\label{Eq:theta_2}
\end{align}
Note that the dynamical system of equations in \cref{Eq:ind_dyn1,Eq:ind_dyn2} describe the time evolution of the respective populations of un-compromised and informed bot devices in the network over time. In order to determine the eventual levels of each type of population in the network, we need to evaluate the equilibrium of the dynamical system. In the subsequent, subsections we focus on analyzing the equilibrium populations of degree $k$ devices.

\vspace{-0.0in}
\subsection{Analysis of Equilibrium State}
 At the equilibrium state, $\frac{d \tilde{B}_k(t)}{dt} = 0$ and $\frac{d BI_k(t)}{dt} = 0$. Therefore, the equilibrium population of degree $k$ un-compromised devices, $\tilde{B}_k^*$ and of informed bot devices, $BI_k^*$ can be expressed as follows:
\begin{align}
&\tilde{B}_k^*(\mu_k) = \frac{\mu_k}{\mu_k + k \sigma_1(\theta^*_{\tilde{B}})}, \label{Eq:eqbm_1}\\
&BI^*_k(\mu_k) = \frac{k^2 \sigma_1(\theta^*_{\tilde{B}}) \sigma_2(\theta^*_{BI})}{(\mu_k + k \sigma_1(\theta^*_{\tilde{B}}))(\beta + \mu_k + k \sigma_2(\theta^*_{BI}))}, \label{Eq:eqbm_2}
\end{align}
with $\theta^*_{\tilde{B}}$ and $\theta^*_{BI}$ denoting the respective probabilities at equilibrium. Note that \cref{Eq:theta_1,Eq:theta_2} expresses $\tilde{B}_k^*$ and $BI_k^*$ in terms of $\theta^*_{\tilde{B}}$ and $\theta^*_{BI}$. However, \cref{Eq:eqbm_1,Eq:eqbm_2} can be used to express $\theta^*_{\tilde{B}}$ and $\theta^*_{BI}$ in terms of $\tilde{B}_k^*$ and $BI_k^*$. Therefore, it presents a self-consistent system of equations which needs to be solved in order to obtain the equilibrium state. An exact solution to the system in analytically challenging. However, an approximate characterization\footnote{Note that these results are based on first order approximation of the first moment of a function of a random variable. Although higher order approximations would lead to tighter approximations, however, it makes the solution analytically complicated precluding subsequent analysis and optimization.} of the probabilities $\theta_{\tilde{B}}$ and $\theta_{BI}$ at equilibrium is provided by the following lemma.
\begin{lemma}\label{equilibrium_lemma}
In a PPP distributed wireless network with D2D communication, the probability of a particular link of a degree $k$ device pointing to an un-compromised and to an informed bot device respectively at equilibrium can be approximately expressed as follows:
\begin{align}
&\theta_{\tilde{B}}^* \approx \min \left(\frac{\mu_k}{\rho \gamma_b p \mathbb{E}[K]} , 1 \right),\\
&\theta_{BI}^* \approx \max \left( 1 - \frac{\mu_k \gamma_c + \rho \gamma_b (\beta + \mu_k)}{\mathbb{E}[K] \rho p \gamma_b \gamma_c} , 0 \right). \label{eqbm_lemma_second_eq}
\end{align}
\begin{proof}
See \textbf{Appendix~\ref{proof_equilibrium_lemma}}.
\end{proof}
\end{lemma}
%Note that the probabilities $\theta_{\tilde{B}}^*$ and $\theta_{BI}^*$ implicitly depend on the degree $k$ of the devices.
These approximations present a lower bound on the actual probabilities. The loss in accuracy for the sake of analytical tractability is discussed in Appendix~\ref{proof_equilibrium_lemma}. Note that Lemma~\ref{equilibrium_lemma} presents an intuitive result where the probability of being connected to an un-compromised device, $\theta_{\tilde{B}}$ is directly proportional to the patching rate and inversely related to the expected degree, vulnerability, malware spreading rate and the transmission success probability. Similar explanation can be derived for $\theta_{BI}$. A direct corollary of the result presented in Lemma~\ref{equilibrium_lemma}, that plays an important role in the optimal patching decisions is provided below:
\begin{corollary} \label{Patching_limit}
For a PPP deployed wireless IoT network being infiltrated by a botnet with malware spreading at a rate $\gamma_b$ and control commands propagating at a rate $\gamma_c$, the upper bound on the required patching rate for a device to have an impact on the equilibrium populations is given~by
\begin{align}
\mu_k \leq  \rho \gamma_b p \mathbb{E}[K] , \ \ \forall k \geq 1,
\end{align}
\begin{proof}
	See \textbf{Appendix~\ref{Patching_limit_proof}}.
\end{proof}
\end{corollary}
This is significant since it provides an estimate of the maximum patching frequency that can be used by the network defender on a degree $k$ device to have an impact on the equilibrium proportions of the devices. In other words, it presents the fundamental limits of the patching rate, since using a higher patching rate than will lead to a completely bot-free population at equilibrium. Similarly, an auxiliary result emanating from~\eqref{eqbm_lemma_second_eq} is expressed in the following Corollary.
\begin{corollary}\label{max_refresh_rate}
The maximum information refresh rate, $\beta$ that can be selected by a bot device to have non-zero informed bot population at equilibrium can be expressed as follows:
\begin{align}
\beta < p \gamma_c \mathbb{E}[K]
\end{align}
\begin{proof}
See \textbf{Appendix~\ref{Patching_limit_proof}}
\end{proof}
\end{corollary}

Although the results presented in Lemma~\ref{equilibrium_lemma} are useful, however, the presence of the minimum and maximum functions present a challenge in leveraging them for optimization purposes. To circumvent this challenge, we propose to use the Log-Sum-Exponential (LSE) function\footnote{The function $\max(x,y)$ can be approximated by $\frac{1}{\eta} \ln \left( e^{\eta x} + e^{\eta y}\right)$ and $\min(x,y)$ can be approximated by $-\frac{1}{\eta} \ln\left( e^{-\eta x} + e^{-\eta y}\right)$ provided that $\eta$ is sufficiently large.} to provide a smooth and continuously differentiable approximation of these expressions. It results in the following:
%\textcolor{blue}{In order to make the probabilities smooth functions of $\mu_k$, we resort to the Log-Sum-Exponential function to convert the minimum to a soft-minimum. The LSE in this case is a concave function of the arguments.}
\begin{align}
&\theta_{\tilde{B}}^* \approx - \frac{1}{\eta} \ln\left( e^{-\eta} + e^{- \eta \left( \frac{\mu_k}{\rho \gamma_b p \mathbb{E}[K]} \right) }   \right),  \label{Eq:theta_b_tilde_lse}\\
&\theta_{BI}^* \approx \frac{1}{\eta} \ln\left( 1 + e^{ \eta \left( 1 - \frac{\mu_k \gamma_c + \rho \gamma_b (\beta + \mu_k)}{\mathbb{E}[K] \rho p \gamma_b \gamma_c} \right) }   \right), \label{Eq:theta_b_i_lse}
\end{align}
where $\eta$ is a sufficiently large constant chosen for accuracy of the \emph{soft-minimum} and \emph{soft-maximum} functions. Note that the LSE relaxation in \cref{Eq:theta_b_tilde_lse,Eq:theta_b_i_lse} may slightly affect the upper bound on the patching rate expressed in Corollary~\ref{Patching_limit} and the upper bound on the possible information refresh rate expressed in Corollary~\ref{max_refresh_rate}. However, the inaccuracy diminishes with the selection of large $\eta$.

\begin{figure*}[t]
	\begin{align}
	%&\tilde{B}_k^* \approx \frac{\mu_{k}}{ \mu_k + \frac{k}{\mathbb{E}[K] }\left(  \rho \gamma_b p \mathbb{E}[K] - \mu_k\right) },  \label{Eq:Final_Eqbm1}\\
	%&BI_k^* \approx \frac{k^2 (\mu_k - \rho \gamma_b p \mathbb{E}[K]) (\mu \gamma_c + \rho \gamma_b (\beta + \mu_k - \rho \gamma_c \mathbb{E}[K]))}{  ( (k - \mathbb{E}[K])\mu_k - k \mathbb{E}[K] \rho \gamma_b p )(  k \mu_k \gamma_c - \gamma_b (  (\beta + \mu_k)(p \mathbb{E}[K] - k \rho ) + k \mathbb{E}[K] \rho \gamma_c p  ))}. \\ \label{Eq:Final_Eqbm2}
	\hspace{-1cm}&\tilde{B}_k^*(\mu_k) \approx  \frac{\mu_k}{\mu_k + k \rho \gamma_b p \left(1 + \frac{1}{\eta} \ln\left( e^{-\eta} + e^{- \eta \left( \frac{\mu_k}{\rho \gamma_b p \mathbb{E}[K]} \right) }   \right)  \right)}, \label{Eq:Final_Eqbm1}\\
	&BI_k^*(\mu_k) \approx \frac{k^2 \rho^2 \gamma_b \gamma_c p \left(1 + \frac{1}{\eta} \ln\left( e^{-\eta} + e^{- \eta \left( \frac{\mu_k}{\rho \gamma_b p \mathbb{E}[K]} \right) }   \right)  \right) }{\left(\mu_k + k \rho \gamma_b p \left(1 + \frac{1}{\eta} \ln\left( e^{-\eta} + e^{- \eta \left( \frac{\mu_k}{\rho \gamma_b p \mathbb{E}[K]} \right) }   \right)  \right) \right)}  \times \frac{\frac{1}{\eta} \ln\left( 1 + e^{ \eta \left( 1 - \frac{\mu_k \gamma_c + \rho \gamma_b (\beta + \mu_k)}{\mathbb{E}[K] \rho p \gamma_b \gamma_c} \right) }   \right)}{\left(\beta + \mu_k + k \rho \gamma_c + \frac{1}{\eta} \ln\left( 1 + e^{ \eta \left( 1 - \frac{\mu_k \gamma_c + \rho \gamma_b (\beta + \mu_k)}{\mathbb{E}[K] \rho p \gamma_b \gamma_c} \right) }   \right) \right)}. \label{Eq:Final_Eqbm2}
	\end{align}
	\hrule
\vspace{-0.2in}
\end{figure*}

Finally, using the results of Lemma~\ref{equilibrium_lemma} and the subsequent LSE relaxation, the equilibrium populations of devices that are un-compromised and devices that are informed bots is expressed by the following theorem:
\begin{theorem}
At equilibrium, the proportion of degree $k$ devices in the network that are un-compromised (not infected with malware), i.e., $\tilde{B}_k^*$ and those that are bots and informed by control commands, i.e., $BI_k^*$ can be approximately expressed by \cref{Eq:Final_Eqbm1,Eq:Final_Eqbm2} respectively.
\begin{proof}
Substitution of \eqref{Eq:theta_b_tilde_lse} into \eqref{Eq:eqbm_1} and \eqref{Eq:theta_b_i_lse} into \eqref{Eq:eqbm_2} leads to this result.
\end{proof}
\end{theorem}

In the following subsection, we make use of the developed analytical model and the approximate results to formulate the network defense problem and subsequently discuss the methodology for solving it.

%$\footnote{For large scale wireless networks, uncorrelation is closer to independence as the mutual degree overlap is relatively small}$.
\vspace{-0.0in}
\subsection{Network Defense Problem \& Solution}
The goal of the network defender is to set up a patching schedule for each network device based on its connectivity in order to prevent the formation of a large scale botnet. The patching rate must take into account the disruption caused to regular operation due to the strategies employed, e.g., firmware upgrade or power cycling, which can be in terms of the downtime of devices. The cost incurred on the operation of a network device due to patching activity is assumed to be a smooth, convex, and increasing function of the patching rate $\mu_k$, represented by $\phi_k :  \mathbb{R}^+ \rightarrow \mathbb{R}^+, \forall k \geq 1$.
%The objective of the defender is to schedule patching of devices in a way so that the system disruption or downtime is minimum subject to a maximum tolerable risk of botnet attacks.
The risk of a botnet formation can be measured in terms of the equilibrium population of devices that are bots and the devices that are receiving control commands assuming knowledge of the transmission rates. Accordingly, targets for the minimum expected proportion of network that is un-compromised and the maximum tolerable proportion of the network that is an informed bot, denoted by $\tau_{\tilde{B}} \in [0,1]$ and $\tau_{BI}\in[0,1]$ respectively, can be set. The network defender's problem can then be formulated as follows:
%The optimal patching rate can be determined by solving the following optimization problem:
\begin{align}
%\begin{aligned}
& \underset{\mu_k, k \geq 1 }{\text{minimize}} & & \hspace{-0cm}\sum_{k=1}^{\infty}  \phi_k ( \mu_k) \pi_k, \label{obj_original}  \\
& \text{subject to} & & \hspace{-0cm} \sum_{k=1}^{\infty}  \tilde{B}_k^*(\mu_k) \pi_k \geq \tau_{\tilde{B}}, \label{const_B_tilde_original}\\
& & & \hspace{-0cm}\sum_{k=1}^{\infty}  BI_k^*(\mu_k) \pi_k \leq \tau_{BI}.  \label{const_BI_original}
%\end{aligned}
\end{align}
The objective represents the total expected cost of patching devices at a rate $\mu_k, \forall k$, while the constraints imply that the average proportion of un-compromised devices in the network must be higher than $\tau_{\tilde{B}}$ and the average proportion of informed bot devices in the network must be smaller than $\tau_{BI}$.
%ensure that the respective population of un-compromised devices and informed bots satisfy the targets.
Note that the constraints are coupled with the objective, which makes the primal problem challenging to solve. Furthermore, despite the fact that the objective is convex, both the constraints may be non-convex in the decision vector since some terms inside the summation are concave while others are convex. This is formally stated in the following lemma.
\begin{lemma} \label{curvature_lemma}
The equilibrium proportion of un-compromised devices, $\tilde{B}_k^*$ is concave in $\mu_k$ for $k < \mathbb{E[K]}$ and convex otherwise. Similarly, there is a change in curvature of the equilibrium proportion of informed bot devices, $BI_k$ from convex to concave with increasing device degree~$k$.
\begin{proof}
See \textbf{Appendix~\ref{Curvature_appendix}}.
\end{proof}
\end{lemma}
Another important observation is that the constraints are linked in terms of the patching rates. A set of patching rates may completely satisfy one of the constraints but not the other. Therefore, it is important to investigate the conditions under which the constraints are active, particularly because there exists a limiting rate at which the constraints saturate. The following lemma presents an important condition relating the target thresholds that determines the status of the constraints.

%It can be shown that the target thresholds play an important part in whether the constraints will be binding or not. It is more formally explained in the following lemma.

\begin{lemma}\label{lemma_max_threshold}
The constraint on the average equilibrium population of informed bots, expressed in~\eqref{const_BI_original}, is always satisfied for any $\tau_{BI} \in [0,1]$ if the target on the average equilibrium population of un-compromised devices is set as follows:
\begin{align}
\tau_{\tilde{B}} \geq \frac{\mathbb{E}[K]p\gamma_c - \beta}{\mathbb{E}[K]p (\rho \gamma_b + \gamma_c)}
\end{align}
\begin{proof}
See \textbf{Appendix~\ref{proof_max_threshold}.}
\end{proof}
\end{lemma}
Therefore, if the condition presented in Lemma~\ref{lemma_max_threshold} is satisfied, we can effectively ignore the constraint~\eqref{const_BI_original} from the optimization problem and proceed with only~\eqref{const_B_tilde_original}. This is extremely important since otherwise, the solution to the optimization problem may be difficult as one of the constraints saturates and is no longer monotonously increasing or decreasing. However, evaluating the condition \emph{a priori}, we can circumvent this difficulty and effectively solve the optimization problem. However, there are several additional challenges. First, since the network is random, there is no upper bound on the maximum possible degree of a device, which makes the optimization problem intractable due to an infinite number of optimization variables. However, due to the structure of the network\footnote{In a PPP network, the probability of having a large number of neighbors decreases faster than the exponential decay rate for sufficiently large degrees.}, it is increasingly rare for a device to have larger degrees. Therefore, we note that there exists a sufficiently large $k = k_{\max}$ such that $\mathbb{P}[K > k_{\max}] \leq \epsilon$, where $\epsilon$ is arbitrarily small. This allows us to convert the optimization problem into one with finite number of optimization variables referred to as $\boldsymbol{\mu} = [\mu_1, \mu_2, \ldots, \mu_{k_{\max}}]^T$. Therefore, the problem can then be expressed as follows:
\begin{align}
\begin{aligned}
& \underset{\boldsymbol{\mu}}{\text{minimize}}
& & \sum_{k=1}^{k_{\max}}  \phi_k ( \mu_k) \pi_k  + \underbrace{\sum_{k = k_{\max}}^{\infty} \phi_k ( \mu_k) \pi_k}_{\epsilon_1} \\
& \text{subject to}
& & \tau_{\tilde{B}} - \sum_{k=1}^{k_{\max}}  \tilde{B}_k^*(\mu_k) \pi_k  - \underbrace{\sum_{k=k_{\max}}^{\infty}  \tilde{B}_k^*(\mu_k) \pi_k}_{\epsilon_2} \leq 0,\\
& & & \sum_{k=1}^{k_{\max}}  BI_k^*(\mu_k) \pi_k + \underbrace{\sum_{k = k_{\max}}^{\infty}  BI_k^*(\mu_k) \pi_k}_{\epsilon_3} -  \tau_{BI} \leq 0.
\end{aligned}
\end{align}
Since, the the Poisson density decays faster than the exponential rate for large degree values, the terms labeled as $\epsilon_1, \epsilon_2$, and $\epsilon_3$ can be made arbitrarily small for sufficiently large $k_{\max}$. Hence, effectively, these terms can be removed and the problem can be converted into a finite optimization problem. Since the primal problem may be non-convex, we resort to solving the dual optimization problem~\cite{dual_methods}. Note, however, that the duality gap in this problem setting is zero and hence solving the dual problem is equivalent to solving the primal problem (See Appendix~\ref{duality_gap} for details). We, therefore, relax the original problem by forming the Lagrangian as follows:
\begin{align}
\mathcal{L}(\boldsymbol{\mu}, \zeta, \xi) &= \sum_{k=1}^{k_{\max}}  \phi_k ( \mu_k) \pi_k  -  \zeta \left(   \sum_{k=1}^{k_{\max}}  \tilde{B}_k^*(\mu_k) \pi_k  - \tau_{\tilde{B}} \right) -  \notag \\ & \qquad \qquad \qquad \qquad \xi \left( \tau_{BI} -  \sum_{k=1}^{k_{\max}}  BI_k^*(\mu_k) \pi_k   \right), \notag \\
&= \sum_{k=1}^{k_{\max}} \left( \phi_k (\mu_k) \pi_k  -  \zeta  \tilde{B}_k^*(\mu_k) \pi_k   + \xi BI_k^*(\mu_k) \pi_k \right) \notag \\ & \qquad \qquad \qquad \qquad \qquad + \zeta \tau_{\tilde{B}} -  \xi \tau_{BI}.
\end{align}
where $\zeta$ and $\xi$ are the Lagrange multipliers, which are dual feasible if $\zeta \geq 0$ and $\xi \geq 0$. The Lagrange dual function can be written as follows:
\begin{align}
\begin{aligned}
g(\zeta, \xi) = \  & \underset{\boldsymbol{\mu} \geq 0}{\min}
   \sum_{k=1}^{k_{\max}}  \left( \phi_k (\mu_k) \pi_k  -  \zeta  \tilde{B}_k^*(\mu_k) \pi_k   + \xi BI_k^*(\mu_k) \pi_k \right) \notag \\ & \hspace{5cm}+ \zeta \tau_{\tilde{B}} -  \xi \tau_{BI}, \notag \\
= \ &  \hspace{-0.0cm} \sum_{k=1}^{k_{\max}}  \Big(  \underset{\mu_k \geq 0}{\min} \big\{ \phi_k (\mu_k) \pi_k  -  \zeta  \tilde{B}_k^*(\mu_k) \pi_k   +   \notag \\ & \hspace{2.5cm} \xi BI_k^*(\mu_k) \pi_k \big\} \Big) + \zeta \tau_{\tilde{B}} -  \xi \tau_{BI}.
\end{aligned}
\end{align}
Note that due to the structure of the Lagrangian, the optimization problem in the dual function decouples in the optimization variables, which makes the complexity of evaluating $g(\zeta, \xi)$ linear in $k_{\max}$~\cite{num_chiang}. For a given pair of Lagrange multipliers, the optimal patching rates $\boldsymbol{\mu}^*$ can be written as follows:
\begin{align}
\mu_k^* = \underset{\mu_k \geq 0}{\arg \min} \big\{ \phi_k (\mu_k) \pi_k  -  \zeta  \tilde{B}_k^*(\mu_k) \pi_k   + \xi BI_k^*(\mu_k) \pi_k \big\}. \label{Mu_opt_interm}
\end{align}
Note that if both $\tilde{B}^*_k$ and $BI_k^*$ are not monotonous in $\mu_k$, it may not be possible to obtain a globally optimal solution for $\mu_k$ in~\eqref{Mu_opt_interm}. However, fortunately using Lemma~\ref{lemma_max_threshold}, we can determine if one of the functions will saturate or not at the optimal $\mu_k$ based on the target thresholds set by the defender. If the condition in Lemma~\ref{lemma_max_threshold} is satisfied, we can ignore the term containing $BI_k^*$ in~\eqref{Mu_opt_interm} and proceed with the optimization\footnote{Removal of the term containing $BI_k^*$ automatically results in the removal of the Lagrange multiplier $\xi$ in the subsequent expressions.}.
Finally, the dual optimization problem can be written as follows:
\begin{equation}
\begin{aligned}
& \underset{\zeta \geq 0, \xi \geq 0}{\text{maximize}}
& &  g(\zeta, \xi) \\
\end{aligned}
\end{equation}
Since $g(\zeta, \xi)$ is a concave optimization problem and has a unique maxima, we can employ a gradient based strategy to achieve the optimal result. However, since a closed form of the dual function may not exist, and hence differentiability may not be guaranteed, we can resort to sub-gradient based iterative update methods for the dual variables~\cite{dual_decomposition}. The sub-gradients of the dual function, evaluated at the optimal patching rates, can be expressed as follows:
\begin{align}
\nabla_{\zeta} g(\zeta, \xi) &= \tau_{\tilde{B}} - \sum_{k=1}^{k_{\max}} \tilde{B}^*_k (\mu_k^*) \pi_k,\\
\nabla_{\xi} g(\zeta, \xi) &= \sum_{k=1}^{k_{\max}} BI^*_k (\mu_k^*) \pi_k - \tau_{BI}.
\end{align}
Therefore, the iterative dual update rule based on the sub-gradients can be expressed as follows:
\begin{align}
\zeta^{(i+1)} &= \Big[\zeta^{(i)} - \alpha \nabla_{\zeta} \Big]^+ ,\notag \\ &=  \Bigg[\zeta^{(i)} - \alpha \left( \tau_{\tilde{B}} - \sum_{k=1}^{k_{\max}} \tilde{B}^*_k (\mu_k^*) \pi_k \right) \Bigg]^+ , \notag \\ & \hspace{4.5cm} i = 0, 1, 2, \ldots, \label{update_zeta} \\
\xi^{(i+1)} &= \Big[\xi^{(i)} - \alpha \nabla_{\xi} \Big]^+, \notag \\  &=  \Bigg[\xi^{(i)} - \alpha \left( \sum_{k=1}^{k_{\max}} BI^*_k (\mu_k^*) \pi_k - \tau_{BI} \right) \Bigg]^+, \notag \\ & \hspace{4.5cm} , i = 0, 1, 2, \ldots, \label{update_xi}
\end{align}
where $\alpha$ is the step size. The complete procedure for obtaining the optimal patching policy is provided in Algorithm~\ref{Alg:Dual}. We initialize the iteration counter $i$ to zero. Furthermore, we initialize the Lagrange multipliers to an arbitrary positive value and set a sufficiently small step-size $\alpha$. Based on the condition $\tau_{\tilde{B}} \geq \frac{\mathbb{E}[K] p \gamma_c - \beta}{\mathbb{E}[K] p (\rho \gamma_b + \gamma_c)}$, we exclude or include the term containing $BI_k^*$ and the associated Lagrange multiplier $\xi$. We then proceed to solve the optimization problem in~\eqref{Mu_opt_interm} for all possible device degrees. Once the optimal intermediate patching rates have been determined, the dual variables are updated based on the sub-gradient based update rule defined in~\eqref{update_zeta} and~\eqref{update_xi}. This process is repeated until the dual variables have converged and the corresponding $\mu_k^*, \forall k = 1, 2, \ldots, k_{\max}$, define the optimal patching rates for each device type. The complete procedure can be shown to have polynomial complexity in the total number of device degrees $k_{\max}$. In the following section, we provide numerical studies to illustrate the behavior of the solutions and its sensitivities with respect to different model parameters.

\begin{algorithm}
\small
\caption{Dual Algorithm to solve the optimal patching problem }\label{Alg:Dual}
\begin{algorithmic}[1]
\Require {Target thresholds, $\tau_{\tilde{B}}$ and $\tau_{BI}$.}
\Initialize{ Iteration $i = 0$, Step-size $\alpha$, Lagrange multipliers $\zeta^{(i)} > 0, \xi^{(i)} > 0$.}
\Repeat
\Procedure{Dual Function Optimization}{}
\If {$\tau_{\tilde{B}} \geq \frac{\mathbb{E}[K] p \gamma_c - \beta}{\mathbb{E}[K] p (\rho \gamma_b + \gamma_c)}$}
\State $\mu_k^{(i)*} \gets \underset{\mu_k \geq 0}{\arg \min} \big\{ \phi_k (\mu_k) \pi_k  -  \zeta  \tilde{B}_k^*(\mu_k) \pi_k\big\}, \  \forall k = 1, \ldots, k_{\max}.$
\Else
\State $\mu_k^{(i)*} \gets \underset{\mu_k \geq 0}{\arg \min} \big\{ \phi_k (\mu_k) \pi_k  -  \zeta  \tilde{B}_k^*(\mu_k) \pi_k   + \xi BI_k^*(\mu_k) \pi_k \big\}, \ \forall k = 1, \ldots, k_{\max}.$
\EndIf
\EndProcedure
\Procedure{Dual Variable Update}{}
\If {$\tau_{\tilde{B}} \geq \frac{\mathbb{E}[K] p \gamma_c - \beta}{\mathbb{E}[K] p (\rho \gamma_b + \gamma_c)}$}
\State $\zeta^{(i+1)} \gets \Bigg[\zeta^{(i)} - \alpha \left( \tau_{\tilde{B}} - \sum_{k=1}^{k_{\max}} \tilde{B}^*_k (\mu_k^*) \pi_k \right) \Bigg]^+$.
\State $\xi^{(i+1)} \gets \Bigg[\xi^{(i)} - \alpha \left( \sum_{k=1}^{k_{\max}} BI^*_k (\mu_k^*) \pi_k - \tau_{BI} \right) \Bigg]^+$.
\Else
\State $\zeta^{(i+1)} \gets \Bigg[\zeta^{(i)} - \alpha \left( \tau_{\tilde{B}} - \sum_{k=1}^{k_{\max}} \tilde{B}^*_k (\mu_k^*) \pi_k \right) \Bigg]^+$.
\EndIf
\EndProcedure
\Until{convergence of $\zeta$ and $\xi$.}
\end{algorithmic}
\end{algorithm}

\vspace{-0.0in}
\section{Results} \label{Sec:Results}
%\textcolor{red}{We pick three test degrees, one around the mean degree, one lower and one higher to illustrate the impact of system parameters on the optimal patching rates.}
In this section, we first describe the network setup and system parameters used for numerical studies. Then, we present the results obtained from the solution to the optimization problem and the associated impact of the parameters involved. The parameters selected for the generation of numerical results are for illustrative purposes and can be modified according to the scenario in practical applications.

Consider a random network of wireless IoT devices distributed according to a homogeneous PPP with intensity $\lambda = 300$ device/km$^2$ and a communication range of $r = 100$ m. On average, a typical IoT device would be able to communicate with $\mathbb{E}[K] = \lambda \pi r^2 = 9.4$, i.e., approximately $9$ other devices.
We assume that the maximum possible degree in the network is $k_{\max} = 25$ for which $\epsilon = \mathbb{P}(K \geq k_{\max})$ is of the order $10^{-6}$. Due to interference and fading effects of the wireless channel during communication, we assume a successful transmission probability of $\rho = 0.95$. We assume that a proportion $p = 0.7$, of the network is vulnerable to be infected by malware. The malware introduced by a botmaster is assumed to transmit packets for infiltration in nearby devices at a rate of $\gamma_b = 0.001$ packets per second (or 1 packet every 1000 seconds) and for control commands propagation at a rate of $\gamma_c = 0.001$ packets per second. The information refresh rate of bots is selected as $\beta = 0.002$ per second. Note that this choice of $\beta$ satisfies the condition provided in Corollary~\ref{max_refresh_rate}.

\begin{figure}
	\centering
	\includegraphics[width=3.2in]{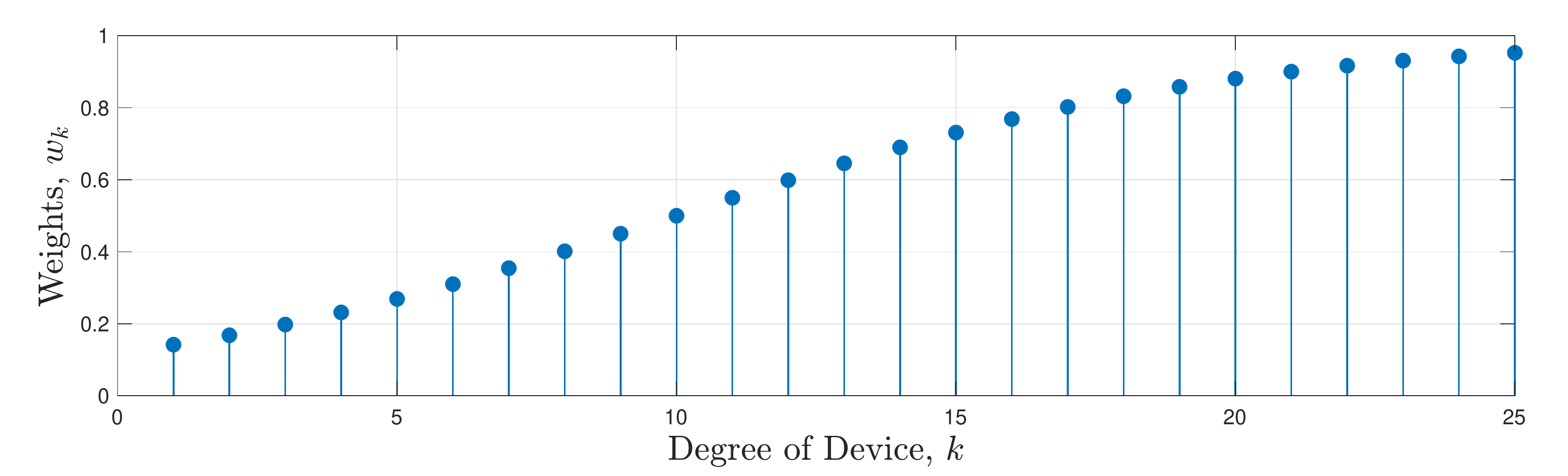}\\
	\caption{Relative impact of unit patching rate of a degree $k$ device on network performance.}\label{Fig:weights}
\end{figure}

In the theoretical analysis, the scaling constant for LSE relaxation of the minimum function is chosen to be $\eta = 100$ for accuracy. The impact of patching a device of degree $k$ on the operational performance of the network is assumed to be captured by the function $\phi_k(\mu_k) = w_k \mu_k^2$, where the weights are modeled using the following logistic function:
\begin{align}
w_k = \frac{1}{1 + e^{-a(k - b)}},
\end{align}
and the constants $a$ and $b$ are chosen to be $a = 0.2$ and $b = 10$ respectively. An illustration of the weight function is provided in Fig.~\ref{Fig:weights}. It implies that a unit patching rate on a device of degree $k$ has a higher impact on network operation as $k$ increases. Hence, it is more costly to increase patching rate for higher degree devices.
%Unless otherwise stated, we assume that a typical device is vulnerable to intrusion attack with a probability $p = 0.7$.
%The information refresh rate $\beta$ is set as $\beta = 0.1$. This signifies the value of recent information by the bots.

\begin{figure*}[t]
\centering
\subfloat[]{\includegraphics[width = 3.2in]{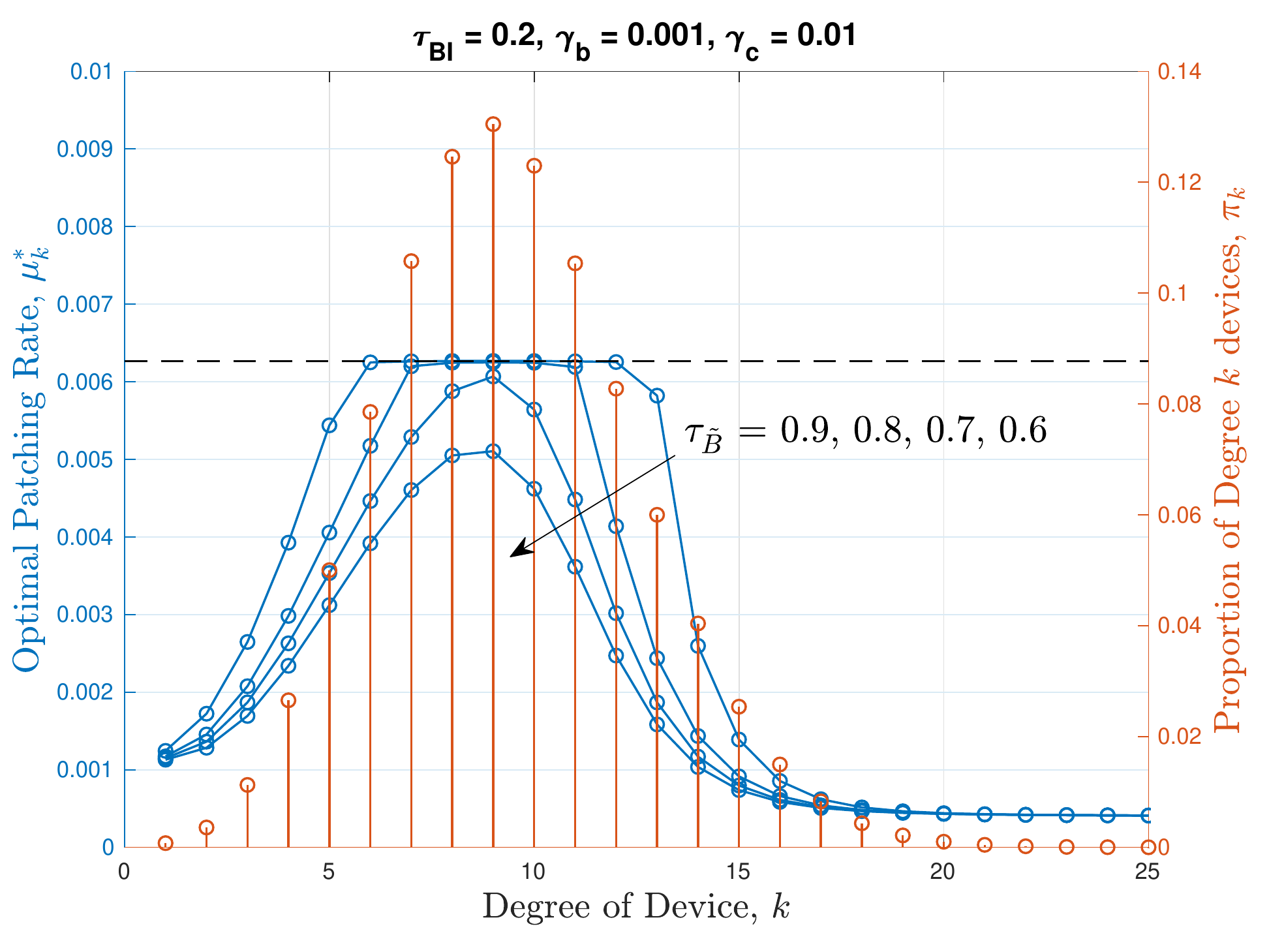} \label{Fig:tau_b_tilde}} \ \
\subfloat[]{\includegraphics[width = 3.2in]{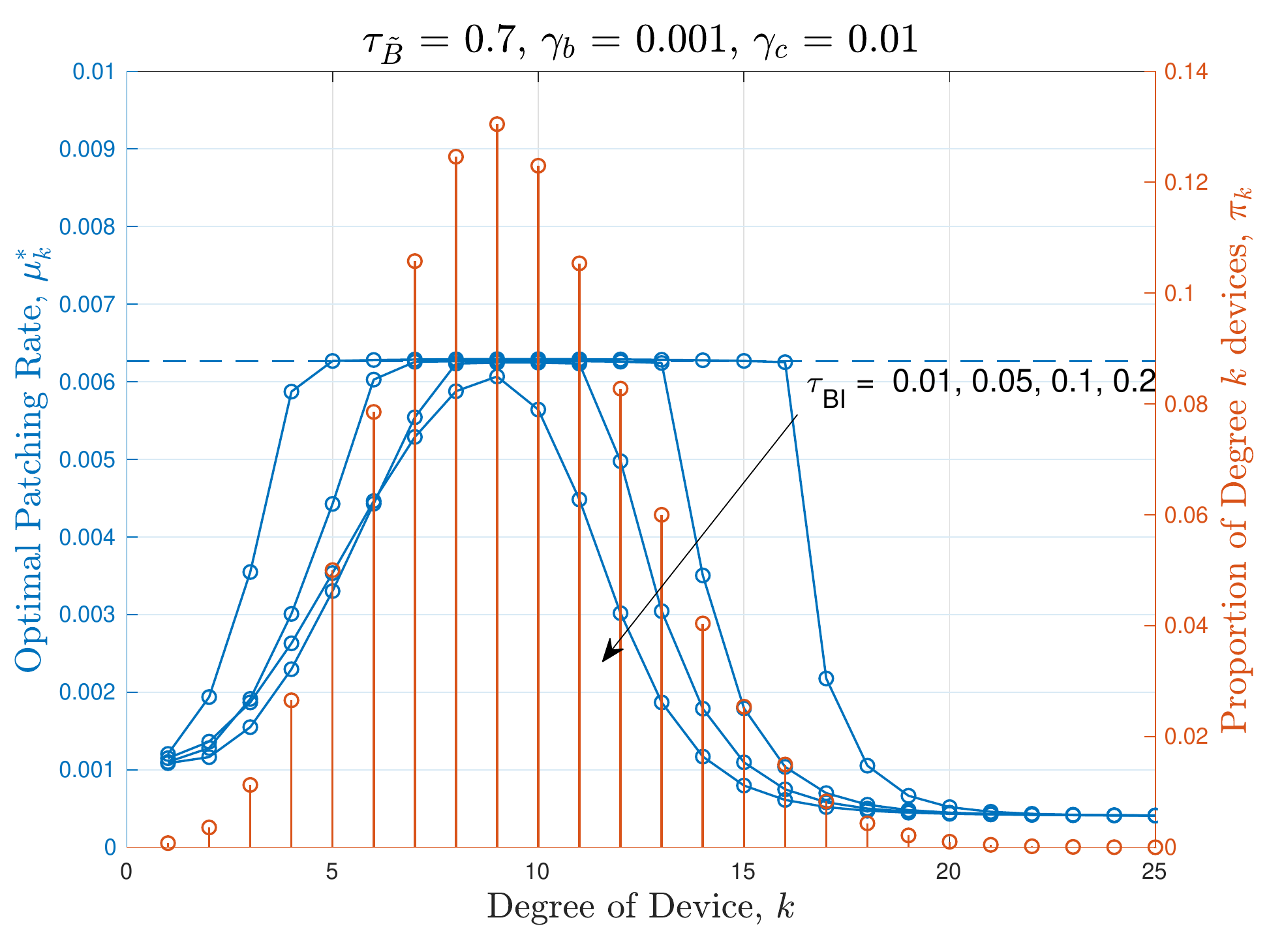} \label{Fig:tau_bi}}
\caption{Impact of varying un-compromised bot proportion threshold $\tau_{\tilde{B}}$ and informed bot proportion threshold $\tau_{BI}$. The dotted line shows the theoretical upper bound expressed in Corollary 1.}
\label{Fig:Optimal_patching_rates}
\end{figure*}

In Fig.~\ref{Fig:tau_b_tilde}, we plot the optimal patching rates for a degree $k$ device in the network with varying target of un-compromised device proportion while fixing $\tau_{BI} = 0.2$. The right axis plots the proportion of degree $k$ devices in the network, or equivalently the probability of a typical device having degree $k$, as a reference for interpreting the results. The dotted line shows the theoretical maximum patching rate that impacts the equilibrium populations as described in Lemma~\ref{Patching_limit}. It can be observed that for $\tau_{\tilde{B}} = 0.6, 0.7$, the optimal patching rates closely follow the proportion of devices due to the monotonously increasing weights $w_k$. However, for more aggressive targets e.g., $\tau_{\tilde{B}} = 0.8, 0.9$, the optimal patching rates saturate for the more probable degrees while increasing patching rates for the lesser probable ones.

In Fig.~\ref{Fig:tau_bi}, we plot the optimal patching rates for a degree $k$ device in the network with varying target of informed bot proportion while fixing $\tau_{\tilde{B}} = 0.7$. Note that a similar behavior is observed in this case where the optimal patching rates closely follow the network degree profile for less aggressive targets, e.g., $\tau_{BI} = 0.1, 0.2$. However, for more aggressive targets such as $\tau_{BI} = 0.01, 0.05$, a saturation is observed for more probable degree types. However, note that the higher and less probable degree devices are patched more frequently although it causes higher disruption since the targets are otherwise not achievable.

\begin{figure*}[t]
\centering
\subfloat[Varying malware spreading rate and bot-free population target.]{\includegraphics[width = 3.2in]{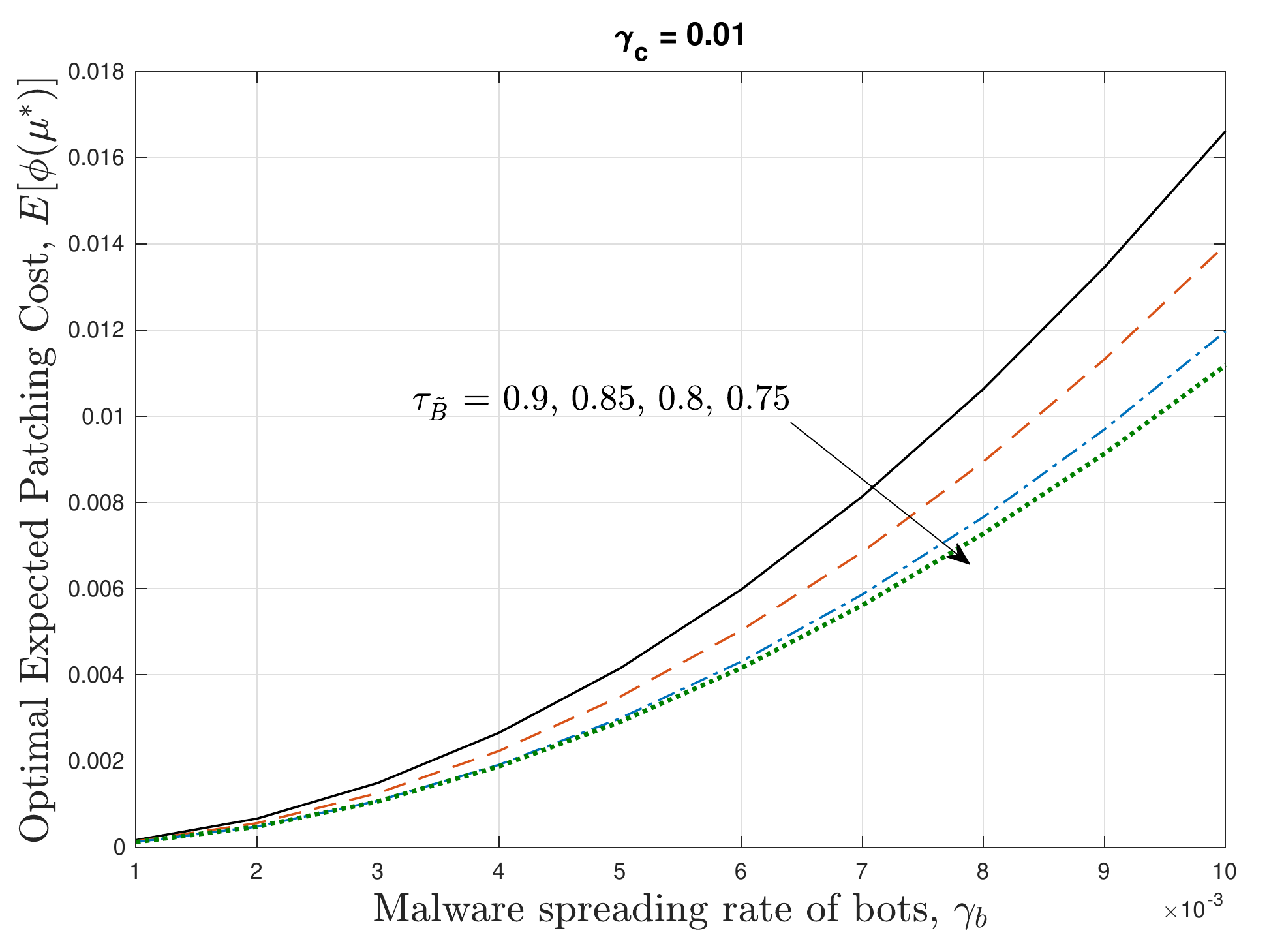} \label{Fig:gamma_b}} \ \
\subfloat[Varying control commands propagation rate and informed bot population target.]{\includegraphics[width = 3.2in]{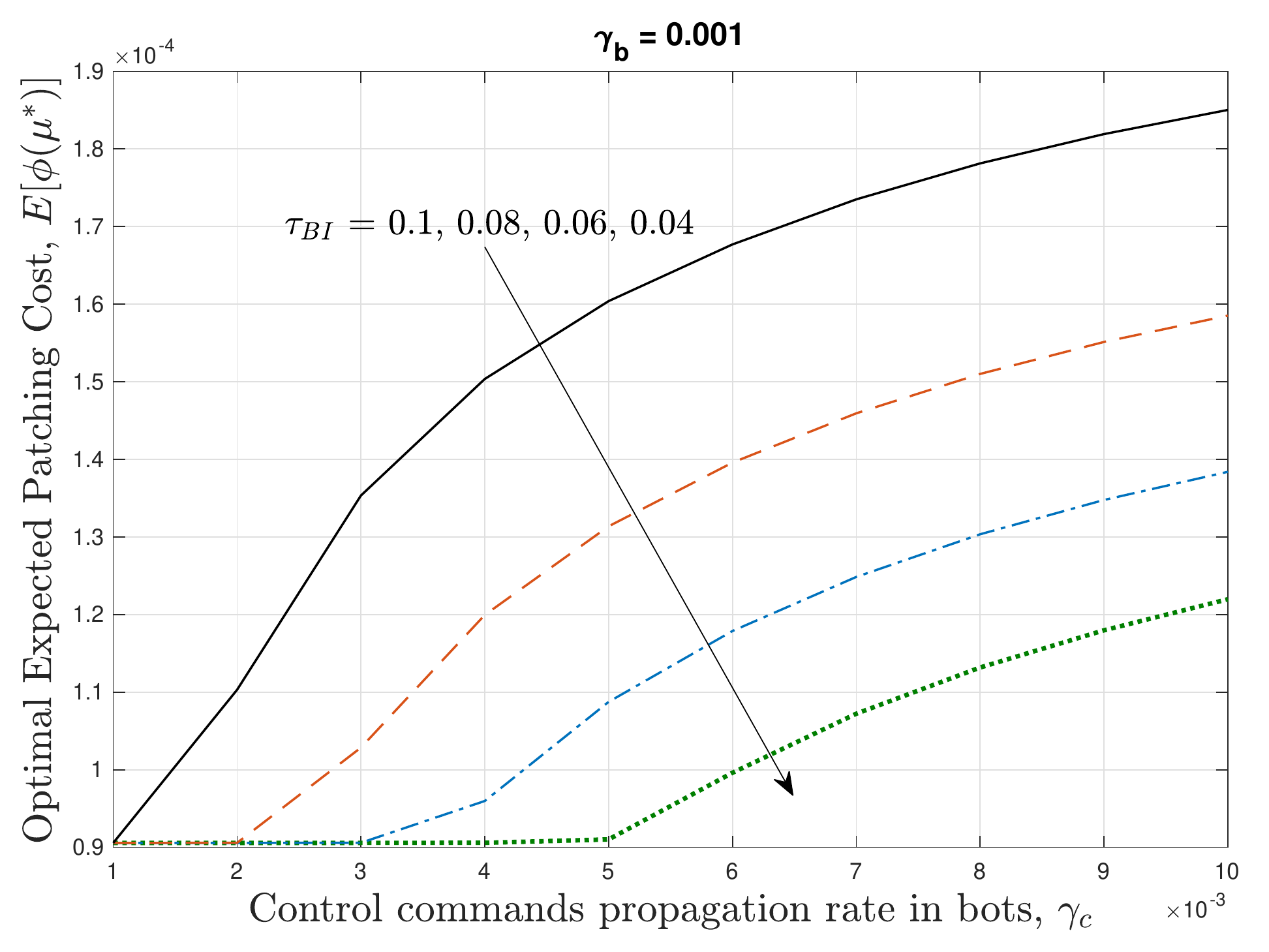} \label{Fig:gamma_c}}
\caption{Expected total cost of patching against varying system parameters.}
\label{Fig:Patching_cost}
\end{figure*}

Finally, Fig.~\ref{Fig:gamma_b} and Fig.~\ref{Fig:gamma_c} illustrate the behavior of the expected total patching cost with varying malware spreading rate and control command spreading rates respectively. It is observed that the expected total patching cost increases at an increasing rate both with increasing malware spreading rate and the target un-compromised device proportion. However,
the expected total patching cost increases at a decreasing rate with increasing control command propagation rate. This shows that the defender is more reactive to the malware spreading rate than the control command propagation rate in terms of a botnet formation. With regards to the effect of varying the device vulnerability in the network as well as the probability of transmission success, a similar behaviour is observed since changing these parameters in turn alters the effective malware propagation rate and the control command propagation rate.

%In this section, we first describe the network parameters before presenting the results. A network of IoT devices uniformly distributed in $\mathbb{R}^2$ according to a PPP of density $\lambda = 1$ is assumed. The devices are assumed to transmit with a power of $P = 1$ and the signal detection threshold is set to achieve an average transmission range of $\bar{R} = 100$ m. For simplicity, we have assumed $p = 1$, i.e., all network nodes are vulnerable to intrusion by the malware. The weights $a$ and $b$ are chosen to be $1$ and $10$ respectively. The relative cost weights for the attacker and defender are selected to be $\xi_1 = 1$ and $\xi_2 = 2$ respectively. This implies that it is cheaper to spread malware in the network as compared to rebooting or patching devices.

%\begin{figure}
%	\centering
%	\includegraphics[width=3in]{Figures/tau_b_tilde.eps}\\
%	\caption{Impact of varying un-compromised bot proportion threshold $\tau_{\tilde{B}}$. The dotted line shows the theoretical upper bound expressed in Corollary 1.\vspace{-0.0in}}\label{Fig:tau_b_tilde}
%\end{figure}

%\begin{figure}
%	\centering
%	\includegraphics[width=3in]{Figures/tau_bi.eps}\\
%	\caption{Impact of varying informed bot proportion threshold $\tau_{BI}$. The dotted line shows the theoretical upper bound expressed in Corollary 1.}\label{Fig:tau_bi}
%\end{figure}

%It can be observed that the optimal patching rates saturate at the upper bound (shown by the dotted line).

%\begin{figure}
%	\centering
%	\includegraphics[width=3in]{Figures/patching_cost_gamma_b.eps}\\
%	\caption{Expected total cost of patching with varying malware spreading rate and bot-free population threshold.\vspace{-0.0in}}\label{Fig:gamma_b}
%\end{figure}
%
%\begin{figure}
%	\centering
%	\includegraphics[width=3in]{Figures/patching_cost_gamma_c.eps}\\
%	\caption{Expected total cost of patching with varying control commands propagation rate and informed bot population threshold.\vspace{-0.0in}}\label{Fig:gamma_c}
%\end{figure}

\begin{figure*}[h]
\centering
\subfloat[]{\includegraphics[width = 2.4in]{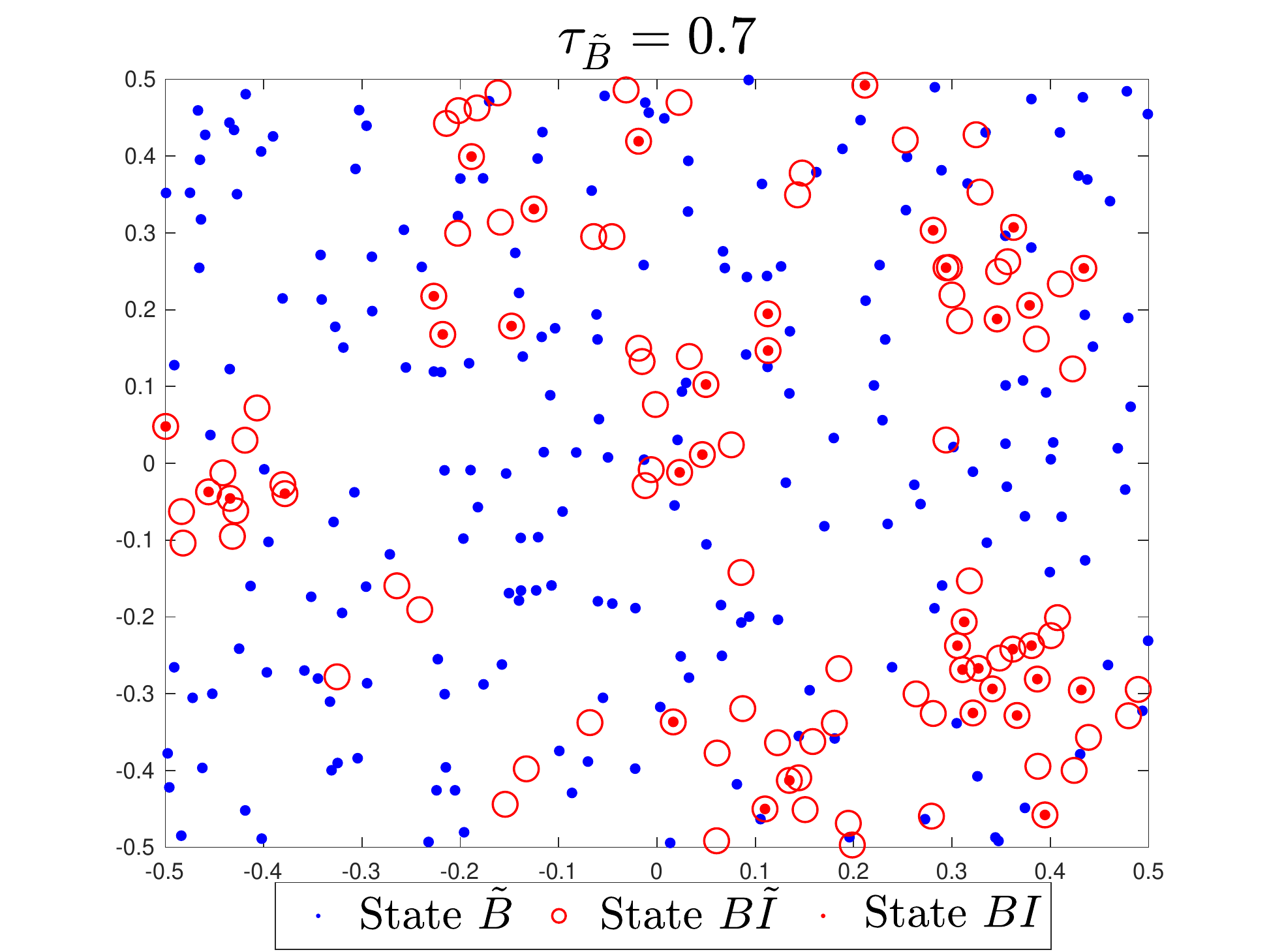} \label{Fig:PPP_snapshot_70}}
\subfloat[]{\includegraphics[width = 2.4in]{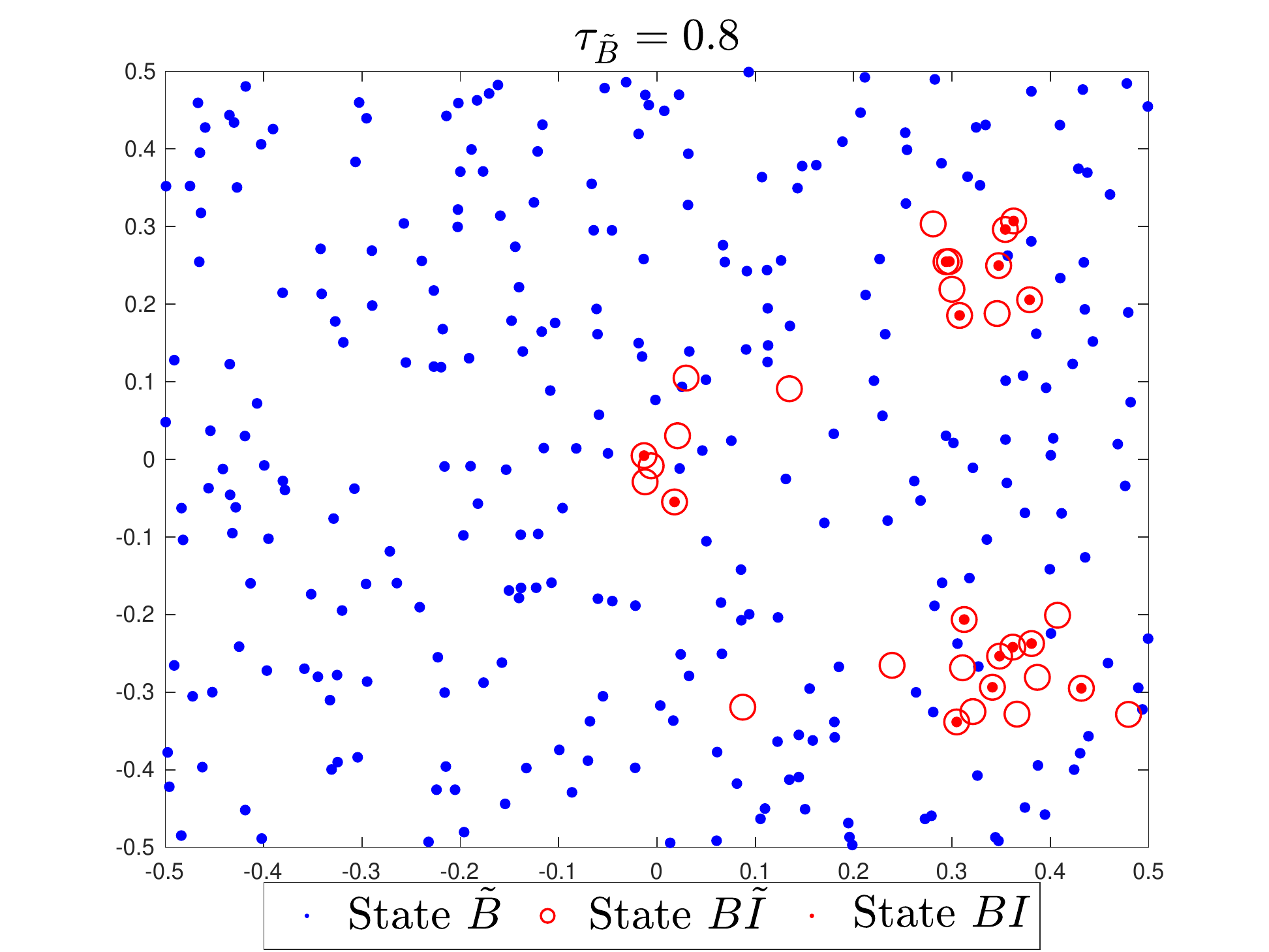} \label{Fig:PPP_snapshot_80}}
\subfloat[]{\includegraphics[width = 2.4in]{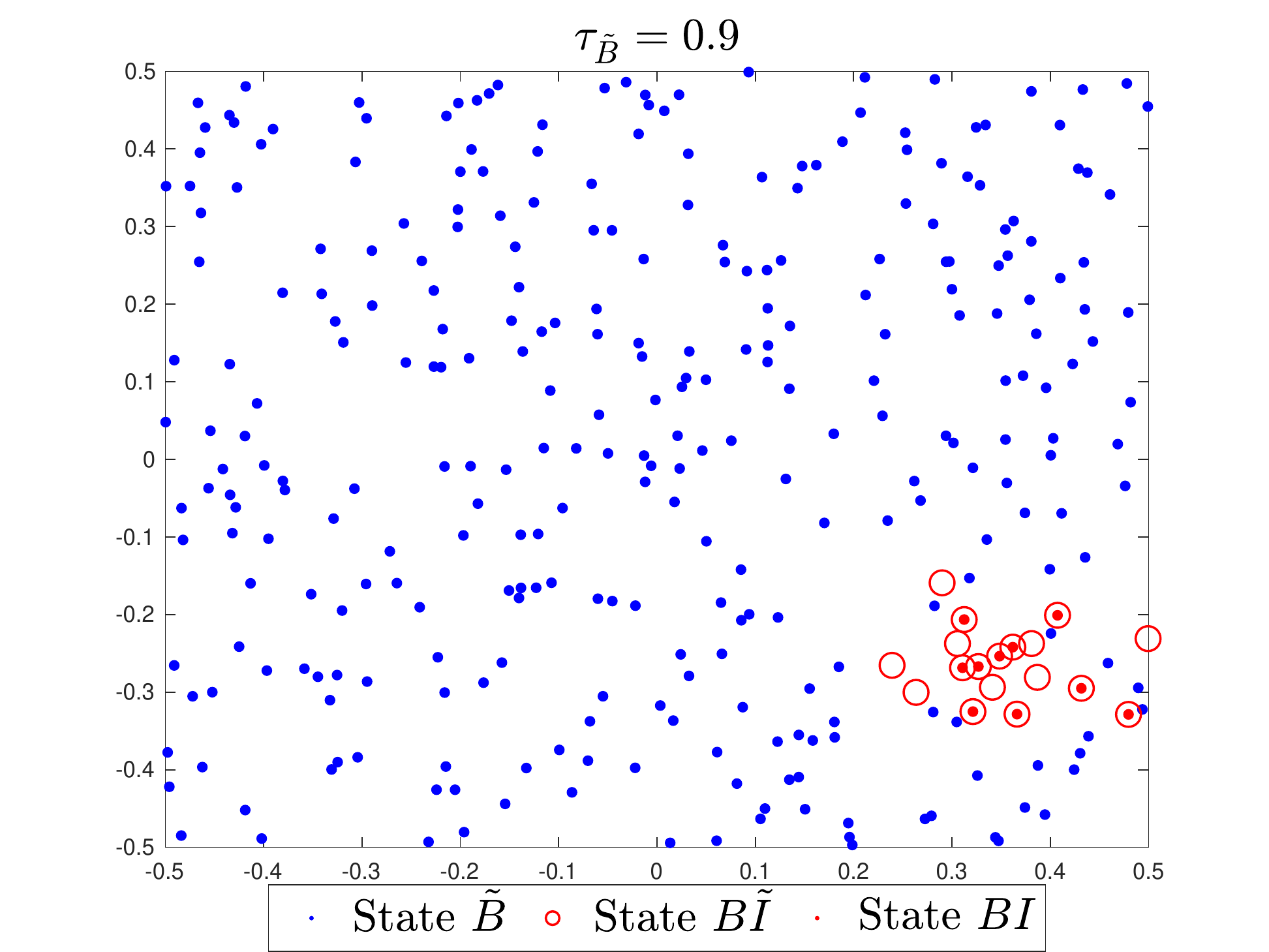} \label{Fig:PPP_snapshot_90}}
\caption{Snapshot of network states at equilibrium in a PPP network.}
\label{Fig:snapshots_PPP}
\end{figure*}

\begin{figure}[h]
	\centering
	\includegraphics[width=3.2in]{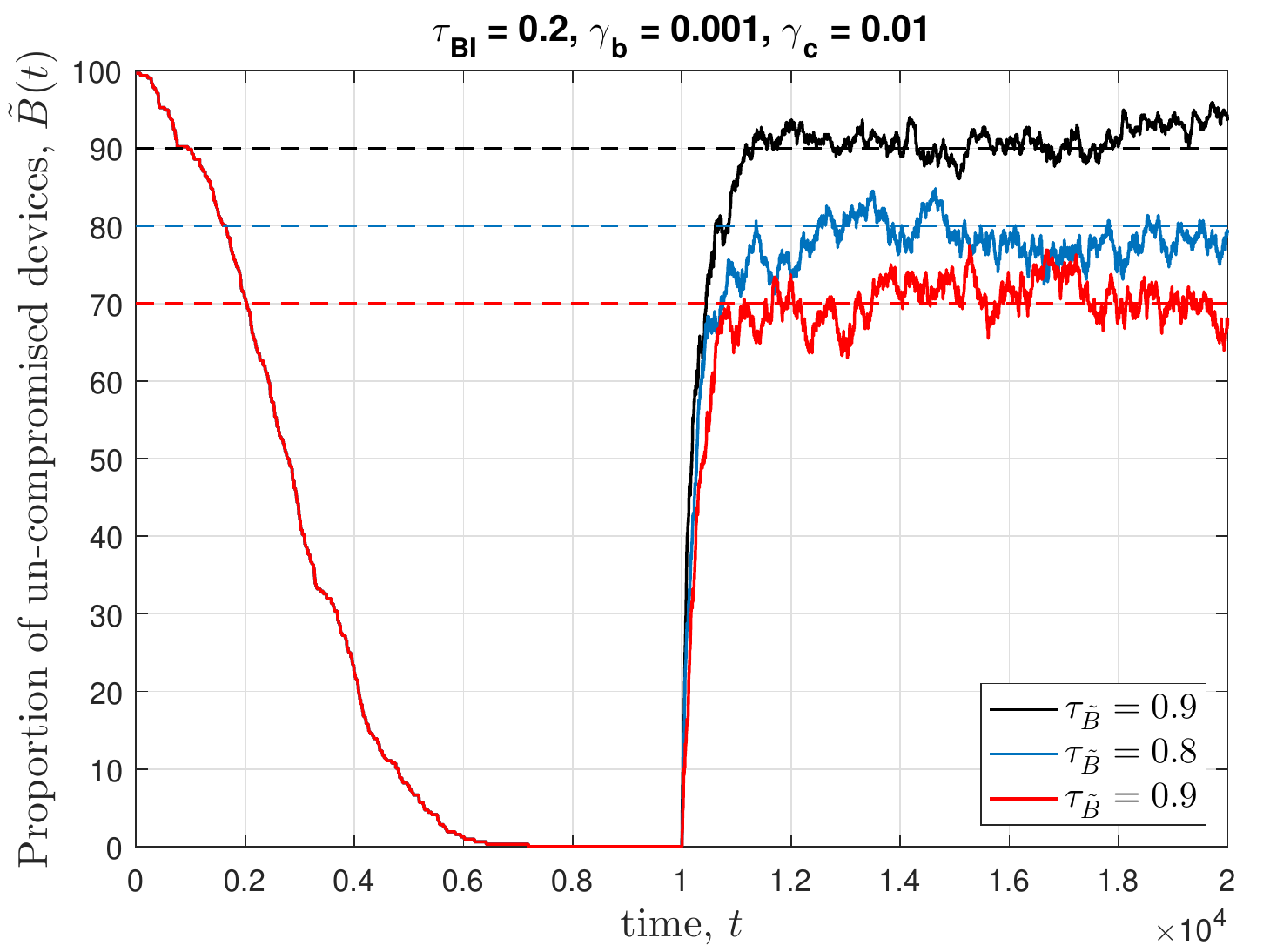}\\
	\caption{Time evolution of the proportion of un-compromised devices in a PPP network.}\label{Fig:PPP_simulation}
\end{figure}

\begin{figure*}[h]
\centering
\subfloat[]{\includegraphics[width = 2.4in]{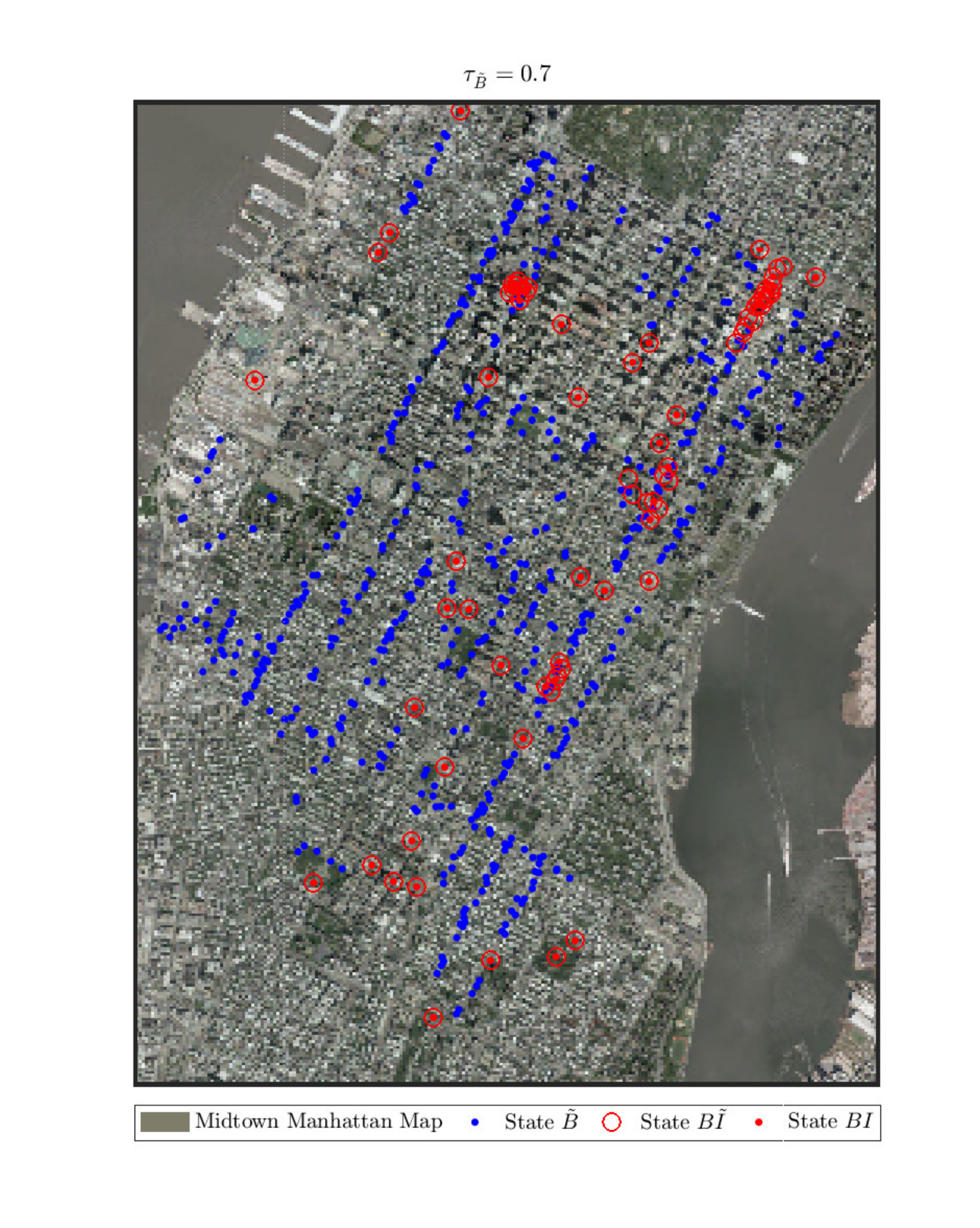} \label{Fig:NYC_snapshot_70}}
\subfloat[]{\includegraphics[width = 2.4in]{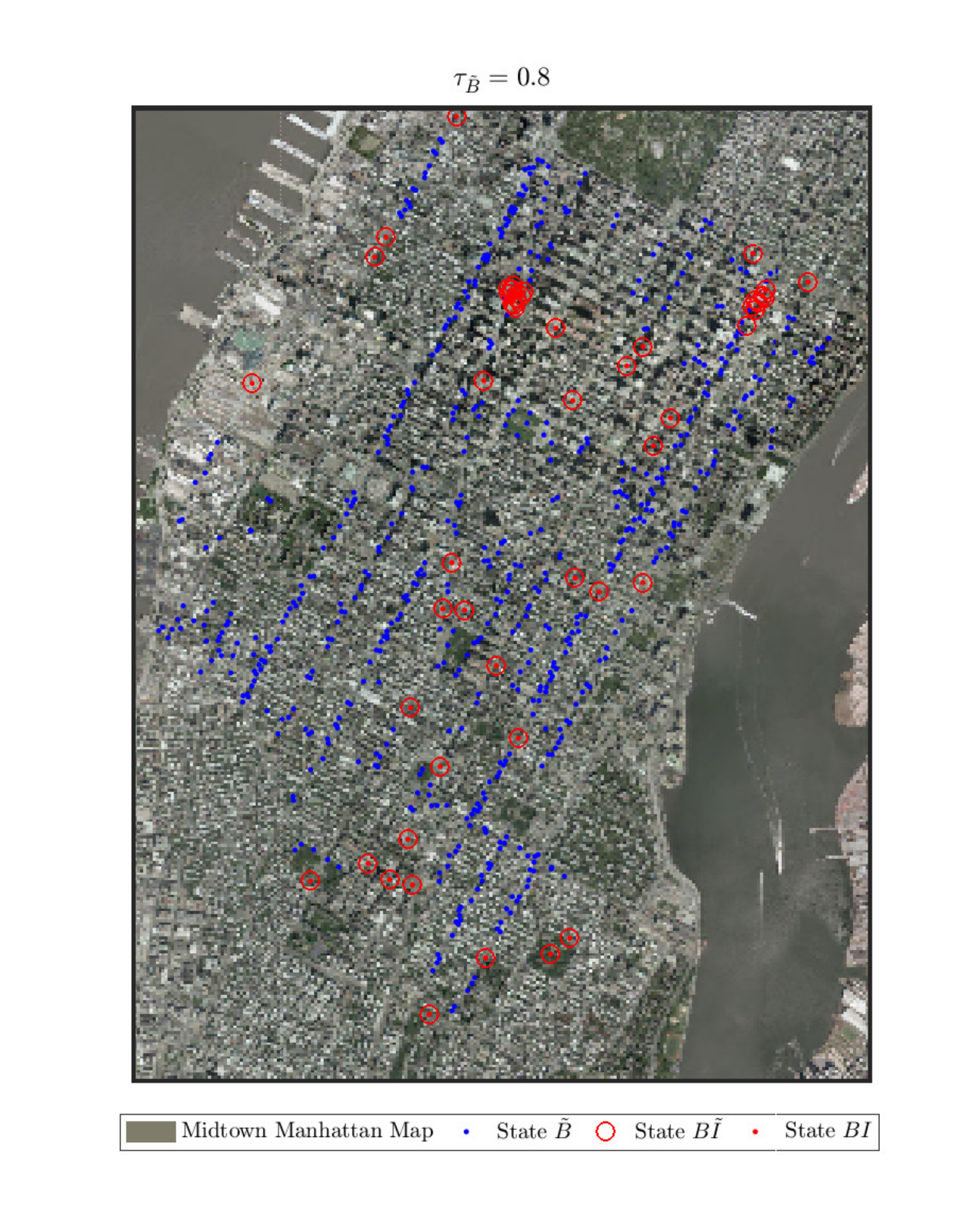} \label{Fig:NYC_snapshot_80}}
\subfloat[]{\includegraphics[width = 2.4in]{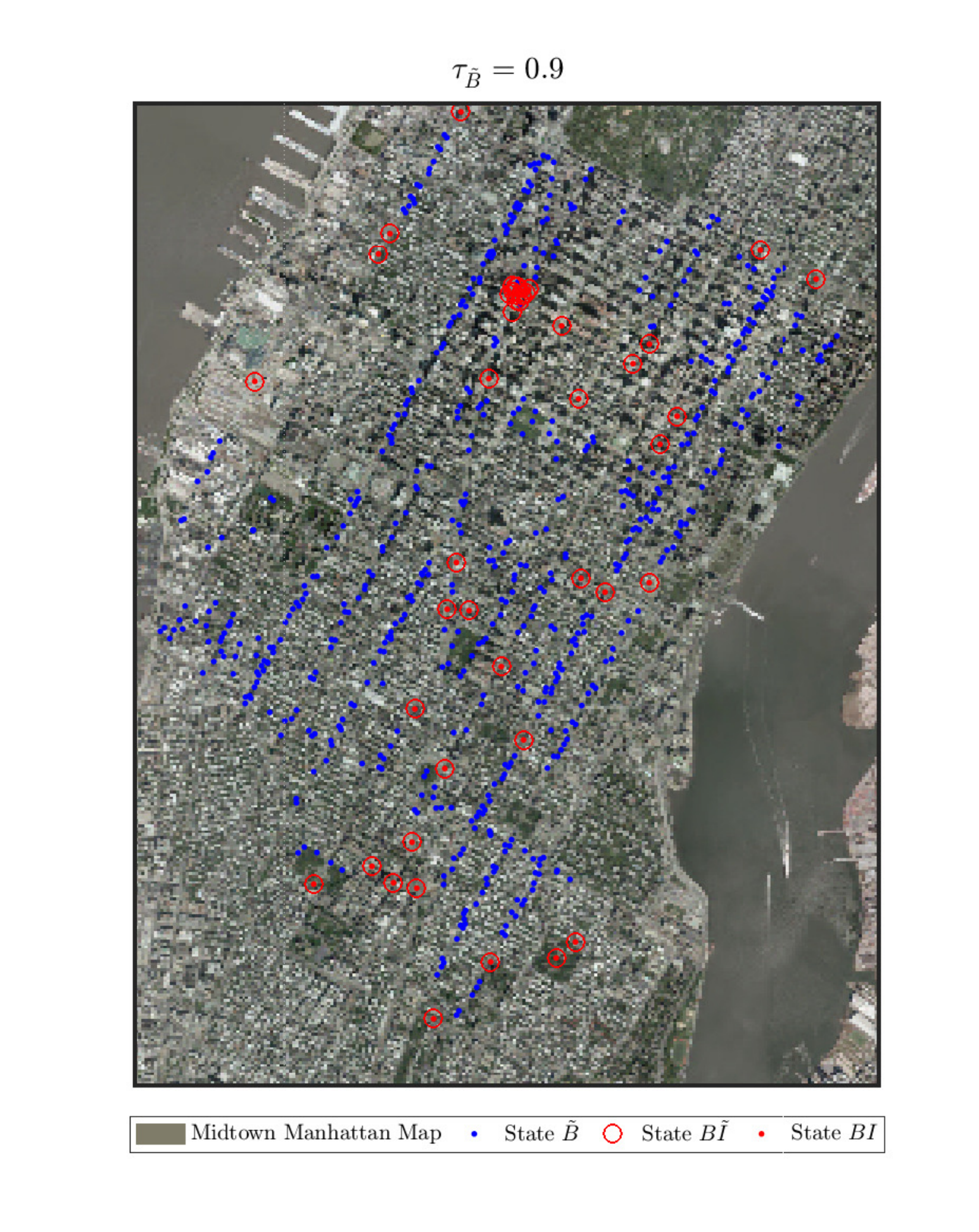} \label{Fig:NYC_snapshot_90}}
\caption{Snapshot of network states at equilibrium in the LinkNYC network.}
\label{Fig:snapshots_NYC}
\end{figure*}

%\begin{figure*}[h]
%\centering
%\subfloat[]{\includegraphics[width = 3.2in]{Figures/PPP_simulation.eps} \label{Fig:PPP_simulation}} \ \
%\subfloat[]{\includegraphics[width = 3.2in]{Figures/NYC_simulation.eps} \label{Fig:LinkNYC_simulation}}
%\caption{Simulation.}
%\label{Fig:Simulation}
%\end{figure*}

\begin{figure}[h]
	\centering
	\includegraphics[width=3.2in]{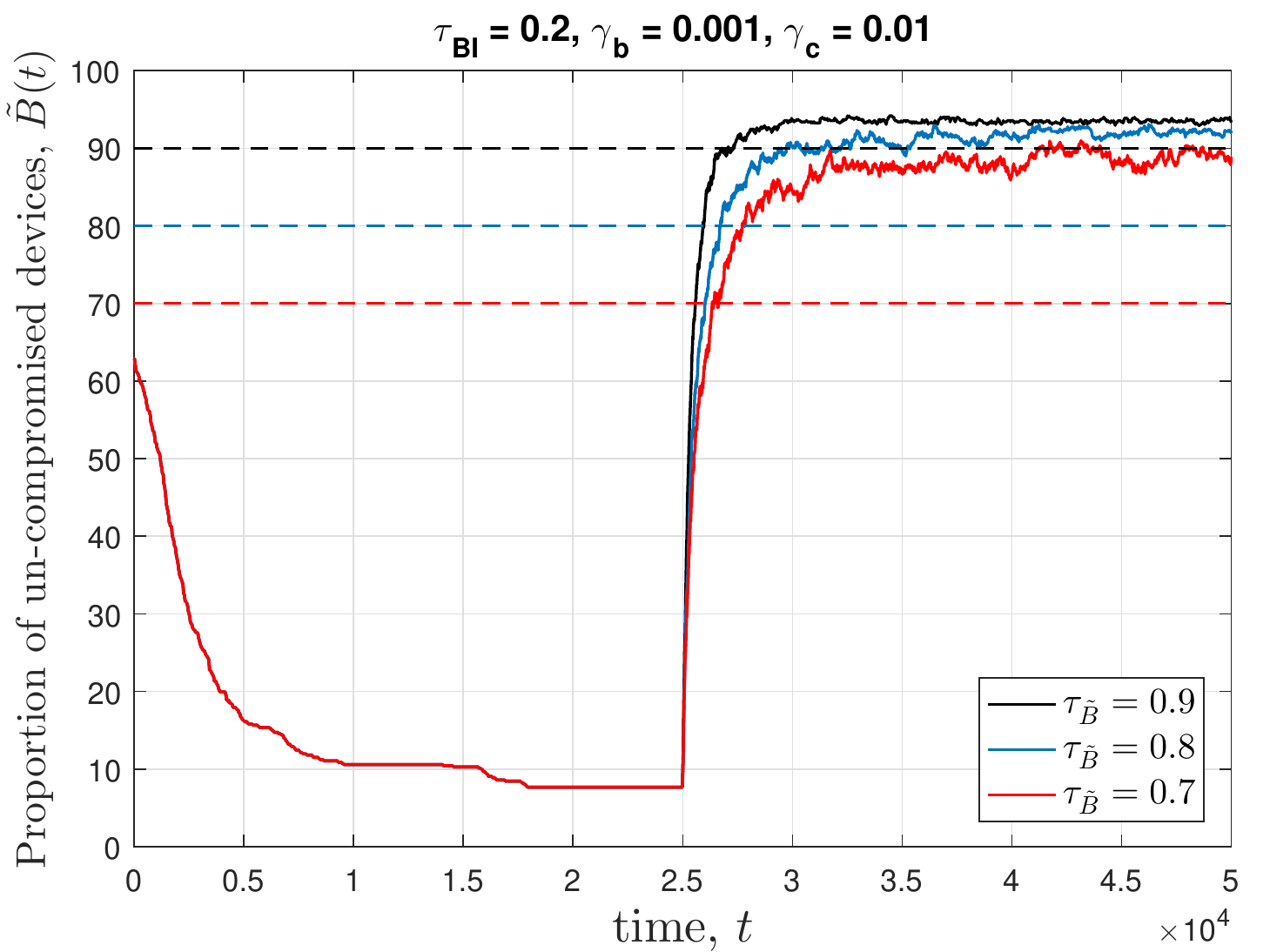}\\
	\caption{Time evolution of the proportion of un-compromised devices in the LinkNYC network.}\label{Fig:LinkNYC_simulation}
\end{figure}

\subsection{Simulation \& Validation}

\textcolor{black}{In this section, we conduct simulation experiments to validate the accuracy of the obtained theoretical results. In the first part, we simulate the considered PPP network. Two different phases are investigated. In the first phase, a malware is introduced at epoch to an arbitrarily selected node and is allowed to propagate to its neighbourhood according to the device vulnerabilities, wireless transmission success probability, as well as the malware propagation rates. The malware spreads from one device to another in a D2D fashion until all the network has been compromised. Note that during the initial phase, there is no patching of devices. During the second phase, the optimal patching policy for each device, based on its degree, is applied on the network. This leads to the recovery of bot devices and the proportions of bots in the network is observed over time. The experiment is repeated for different target thresholds for bot-free population, i.e., $\tau_{\tilde{B}} = 0.7, 0.8$ and 0.9. Fig.~\ref{Fig:snapshots_PPP} illustrates a snapshot of the device states in the network after reaching equilibrium. Note that more devices are un-compromised at equilibrium as $\tau_{\tilde{B}}$ increases as reflected by Fig.~\ref{Fig:PPP_snapshot_70},~\ref{Fig:PPP_snapshot_80}, and~\ref{Fig:PPP_snapshot_90}. The time evolution of un-compromised devices for each of the thresholds is recorded in Fig.~\ref{Fig:PPP_simulation}. Notice that the proportion of un-compromised devices increasingly drops from 100\% to 0\% as the malware is allowed to propagate in the network. However, when the patching process is started in the second phase (i.e., $t = 10^4$), the bot-free population sharply rises until it reaches the target threshold. Although the population keeps fluctuating due to the ongoing dynamical processes but on average the policy is observed to accurately achieve the defined targets. \newline
\indent To further illustrate the usefulness and impact of our proposed methodology and obtained results, we simulate an experiment on the actual LinkNYC hotspot locations data. We assume that IoT devices are placed at each of these locations with a communication range of 140 m. Again, the simulation is carried out in two phases. In the first phase, the malware is allowed to propagate in the network until it has achieved the maximum spread. To ensure complete penetration of the malware in the network, we initially introduce the malware in nodes which have a degree of 2. This allows the propagation of the malware from one device to another over time until it affects most of the nodes during the first phase. Note that this network is not exactly a PPP, the malware spread is not as effective since some nodes may be isolated or clustered together. Similarly, during the second phase (i.e., $t = 2.5 \times 10^4$), the patching process is started until the equilibrium is achieved. Again, the experiment is repeated for different target thresholds for bot-free population, i.e., $\tau_{\tilde{B}} = 0.7, 0.8$ and 0.9. The snapshots of the network states at equilibrium are shown in Fig.~\ref{Fig:snapshots_NYC}. A similar behaviour is observed as the network increasingly becomes bot free at equilibrium as the patching rates are increased. The time evolution of un-compromised devices for each of the thresholds is recorded in Fig.~\ref{Fig:LinkNYC_simulation}. We start off with infecting around 40\% of the devices with malware and allow it to spread. It results in an infection of around 92\% of the network with 8\% un-compromised devices. However, once the patching policy is implemented, the network recovers sharply and is able to achieve much higher bot-free proportions than the target. It is pertinent to mention that since the network is not a PPP, the spread of malware is more difficult. Hence, the developed patching policy is more effective than expected, resulting in better performance of the policy. Therefore, a Poisson network assumption proves to be a more conservative approximation of the real network, which is favourable in practice as the results correspond to a worst case scenario.
}

\vspace{-0.0in}
\section{Conclusion \& Future Work} \label{Sec:Conclusion}
In this paper, we develop a mathematical model to study the formation of botnets in wireless IoT networks. A customized dynamic population process model coupled with a Poisson point process based network model is proposed to capture the evolution of different types of population in the network while keeping the network geometry into account. The proposed model characterizes the behaviour of malware transmission from one device to another using the wireless interface along with the propagation of control commands between bot devices in the network. A network defender is assumed to patch the devices to avert the formation of a botnet that may trigger a coordinated attack at a later stage. The equilibrium state of malware infection and message propagation in the devices is determined using approximate analysis. The results are then used to develop a network defense problem that aims to obtain optimal patching rates while minimizing the disruption to regular network operation under tolerable botnet activity. While the optimal patching problem may be non-convex, a dual decomposition algorithm with appropriate conditions is proposed to solve the optimization problem resulting in the optimal patching schedule for network devices based on their connectivity profile.

%As part of the extension of this work, I plan to capture the DDoS attack risk and present it as a key metric in the paper.
In this work, the network defender's problem has been studied based on the knowledge of the attacker behavior and strategies. However, the defender's actions may also impact the attacker's strategies. Therefore, as part of the future work, we intend to use the proposed model as a basis for developing a game theoretic framework which will enable us to derive optimal policies for both the attacker and defender.

%This metric can be used to obtain the critical patching or recovery rates in the network. Some key concepts such as random patching and targeted patching can be key contributions to the model. In my opinion, using a patching rate proportional to the degree of the devices may result in simpler and convenient equilibrium expressions. However, its investigation is left for the complete version of this work. Moreover, the results can reflect the attacker and defender costs. Several other results can also be presented such as the equilibrium bot propagation etc.

\appendices

\vspace{-0.0in}
\section{Proof of Lemma~\ref{equilibrium_lemma}} \label{proof_equilibrium_lemma}
\begin{figure*}[t]
\centering
\subfloat[]{\includegraphics[width = 3.1in]{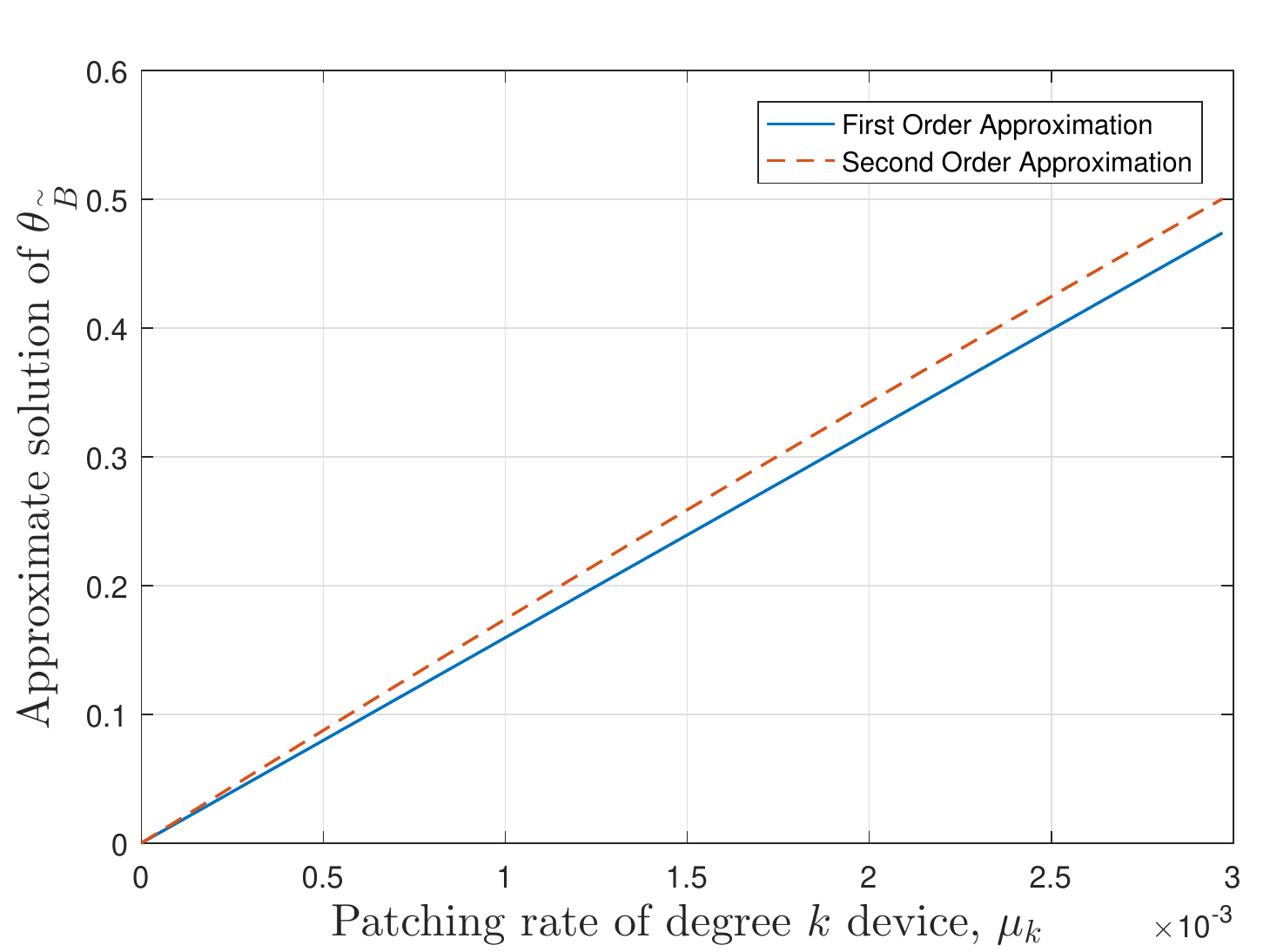} \label{Fig:approx_theta_b_tilde}} \ \
\subfloat[]{\includegraphics[width = 3.1in]{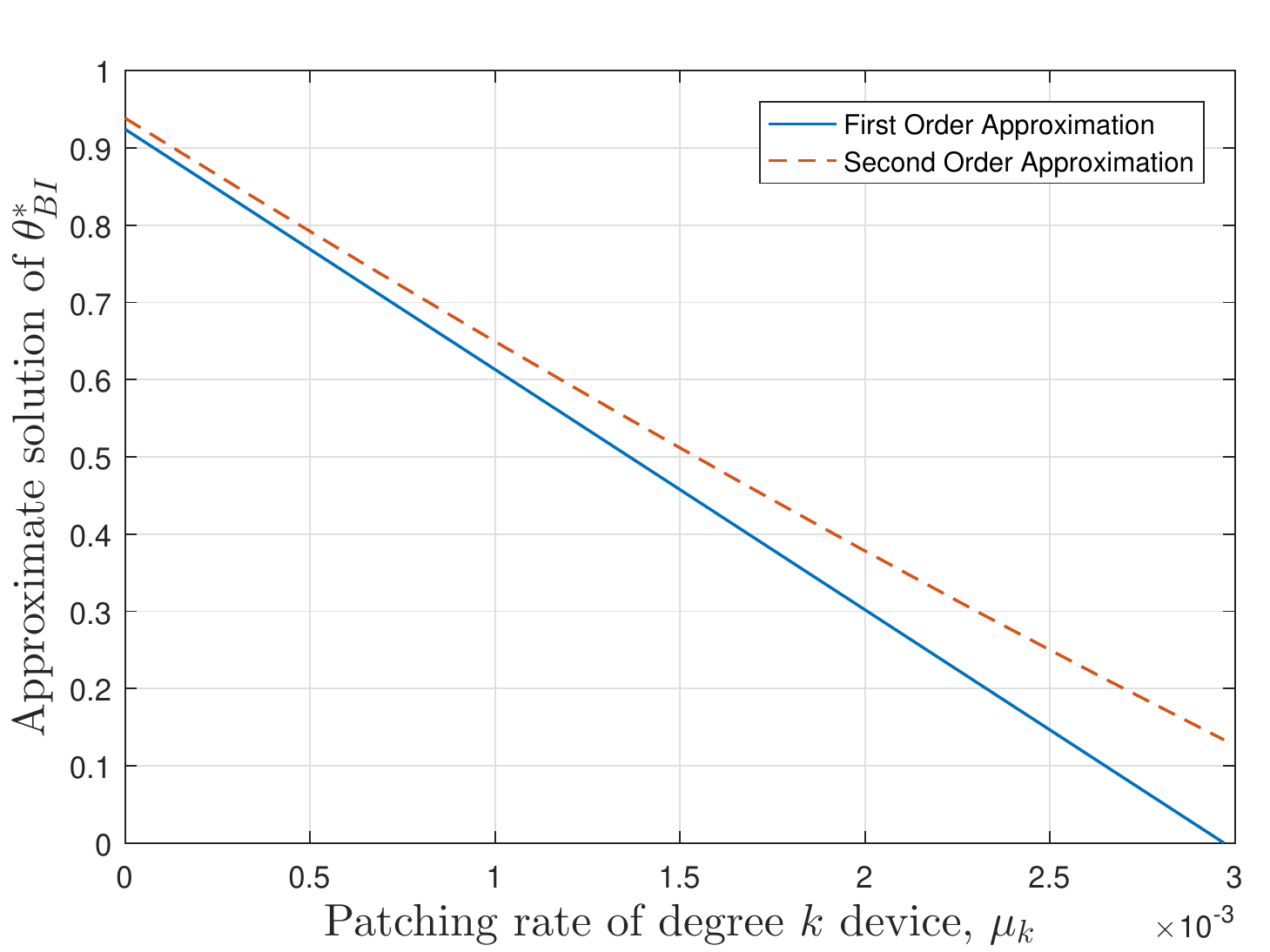} \label{Fig:approx_theta_bi}}
\caption{Approximation accuracy of link probabilities.}
\label{Fig:approx_accuracy}
\end{figure*}

\noindent By substituting~\eqref{Eq:eqbm_1} into~\eqref{Eq:theta_1}, we arrive at the following equation that needs to be solved for $\theta_{\tilde{B}}$:
\begin{align} \label{Eq:theta_b_tilde_self}
\theta_{\tilde{B}} &= \sum_{k^{\prime}} \frac{k^{\prime} P(k^{\prime})}{\mathbb{E}[K]} \left( \frac{\mu_{k^{\prime}}}{\mu_{k^{\prime}} + k^{\prime} \sigma_1(\theta_{\tilde{B}})} \right), \notag \\ & = \sum_{k^{\prime}} \frac{k^{\prime} P(k^{\prime})}{\mathbb{E}[K]} \left( \frac{\mu_{k^{\prime}}}{\mu_{k^{\prime}} + k^{\prime} \rho \gamma_b p (1 - \theta_{\tilde{B}})} \right).
\end{align}
The optimal $\theta_{\tilde{B}}$ is referred to as $\theta_{\tilde{B}}^*$.
The first step is to make use of the degree independence in a homogeneous PPP network to write~\eqref{Eq:theta_b_tilde_self} as follows:
\begin{align} \label{Eq:theta_b_tilde_expect}
&\theta_{\tilde{B}}^* = \mathbb{E} \Bigg[ \frac{\mu_k}{\mu_{k} + k \rho \gamma_b p (1 - \theta_{\tilde{B}}^*)} \Bigg].
\end{align}
Due to the complex form of $\mathbb{P}(K=k)$, a tractable closed form for $\mathbb{E} \big[ \frac{\mu_k}{\mu_{k} + k \rho \gamma_b p (1 - \theta_{\tilde{B}}^*)} \big]$ cannot be easily obtained. Using Taylor expansions for the moments of functions of random variables, the expectation of a function $g(.)$ can be expressed as $\mathbb{E}[g(K)] \approx g(\mathbb{E}[K]) + \frac{g^{\prime \prime}(\mathbb{E}[K])}{2} \sigma_K^2$, where $\sigma_K$ is the variance of the degree. However, using a second order approximation results in loss of tractable solution for~\eqref{Eq:theta_b_tilde_expect}. Therefore, we resort to the first order approximation for simplicity, which results in~\eqref{Eq:theta_b_tilde_expect} being expressed as follows:
\begin{align}
\theta_{\tilde{B}}^* \approx  \frac{\mu_k}{\mu_{k} + \mathbb{E}[K] \rho \gamma_b p (1 - \theta_{\tilde{B}}^*)},
\end{align}
It can be solved for $\theta_{\tilde{B}}^*$ to lead to the following:
\begin{align}
\theta_{\tilde{B}}^* \approx \frac{\mu_k}{\rho \gamma_b p \mathbb{E}[K]}.
\end{align}
Note that since $\mu_k \geq 0$ is not bounded from above, so $\theta_{\tilde{B}}^*$ may become higher than unity which is not possible since it represents a probability. Therefore, we restrict it from above by unity, thus proving the first part of the lemma.
Using a similar methodology, substituting~\eqref{Eq:eqbm_2} into~\eqref{Eq:theta_2} leads to the following expression for $\theta_{BI}^*$:
\begin{align}
\theta_{BI}^* &= \mathbb{E} \Bigg[ \frac{k^2 \sigma_1(\theta_{\tilde{B}}) \sigma_2(\theta_{BI}^*)}{(\mu_k + k \sigma_1(\theta_{\tilde{B}}))(\beta + \mu_k + k \sigma_2(\theta_{BI}^*))} \Bigg], \notag \\ & = \mathbb{E} \Bigg[ \frac{k^2 \sigma_1(\theta_{\tilde{B}}) \rho \gamma_c \theta_{BI}^*}{(\mu_k + k \sigma_1(\theta_{\tilde{B}}))(\beta + \mu_k + k \rho \gamma_c \theta_{BI}^*)} \Bigg].
\end{align}
Again, using the first order approximation of the function inside the expectation, we arrive at solving the following equation:
\begin{align}
\theta_{BI}^* \approx   \frac{(\mathbb{E}[K])^2 \sigma_1(\theta_{\tilde{B}}) \sigma_2(\theta_{BI}^*)}{(\mu_k + \mathbb{E}[K] \sigma_1(\theta_{\tilde{B}}))(\beta + \mu_k + \mathbb{E}[K] \sigma_2(\theta_{BI}^*))} , \notag \\
\end{align}
Solving this for $\theta_{BI}^*$, after some algebraic manipulations, leads to the following result:
\begin{align}
\theta_{BI}^* \approx 1 - \frac{\mu_k \gamma_c + \rho \gamma_b (\beta + \mu_k)}{k \rho p \gamma_b \gamma_c}.
%\theta_{BI}^* \approx \frac{ \mu_k ( \beta + \mu_k) + k \rho \gamma_b p (\beta + \mu_k - k \rho \gamma_c)(1 - \theta_{\tilde{B}}^*)}{ - k \rho \gamma_c \left(\mu_k + k \rho \gamma_b p (1 - \theta_{\tilde{B}}) \right) }
\end{align}
Since $\mu_k$ represents a probability, it needs to be non-negative. Hence, $\theta_{BI}^*$ needs to be restricted at 0 from below, leading to the result provided in Lemma~\ref{equilibrium_lemma}. In Fig.~\ref{Fig:approx_accuracy}, we plot the results obtained from the first order and second order approximations of the probabilities $\theta^*_{\tilde{B}}$ and $\theta^*_{BI}$ against the patching rates. It is observed that the gap between the approximations increases as the patching rate gets higher. Furthermore, the approximations for $\theta^*_{\tilde{B}}$ are relatively much closer as compared to the ones for $\theta^*_{BI}$. Therefore, despite some loss in accuracy, it is still reasonable to use the first order approximations due to the powerful analytical tractability, that facilitates further analysis and decision-making.
%\section{Proof of Lemma~\ref{rfp_dist}} \label{proof_rfp_dist}
%The proof can be done using the cumulative distribution function (\emph{cdf}) approach. The \emph{cdf} of the random relay distance can be expressed as $F_{\mathbf{r}}(r) = \mathbb{P}[\mathbf{r} \leq r] =\pi r^{2}/ \pi \bar{R}^{2}$. Differentiating the \emph{cdf} with respect to $r$ gives $f_{\mathbf{r}}(r)$ as shown in Lemma~\ref{rfp_dist}.

\vspace{-0.0in}
\section{Proof of Corollary~\ref{Patching_limit}} \label{Patching_limit_proof}
From \eqref{Eq:eqbm_1}, we deduce that in order for $\theta_{\tilde{B}}^*$ to assume  a nontrivial value, $\frac{\mu_k}{\rho \gamma_b p \mathbb{E}[K]}$ must be smaller than unity. This implies that $\mu_k \leq \rho \gamma_b p \mathbb{E}[K]$. Similarly, from \eqref{Eq:eqbm_2}, we deduce that $\frac{\mu_k \gamma_c + \rho \gamma_b (\beta + \mu_k)}{\mathbb{E}[K] \rho p \gamma_b \gamma_c} \leq 1$ in order for $\theta_{BI}^*$ to assume a non-trivial value. It results in the condition $\mu_k \leq \frac{\rho \gamma_b \gamma_c p \mathbb{E}[K] - \rho \gamma_b \beta}{\gamma_c + \rho \gamma_b}$ with an implicit condition $\beta < p \gamma_c \mathbb{E}[K]$ for it to be meaningful. It is formally expressed as Corollary~\ref{max_refresh_rate}. However, the upper bound obtained from \eqref{Eq:eqbm_1} is higher, thus becoming the effective upper bound. Therefore, any $\mu_k$ higher than the upper bound is futile in having an impact on the equilibrium state of the devices. In other words, patching devices at a higher rate than the upper bound only affects the regular network operation without having any impact on botnet formation.

%\textcolor{red}{For feasible $\mu_k$, we illustrate the curvature of the probabilities in Fig.}

\section{Proof of Lemma~\ref{curvature_lemma}} \label{Curvature_appendix}
\begin{figure*}[t]
\centering
\subfloat[Equilibrium populations against patching rate for small degree devices.]{\includegraphics[width = 3in]{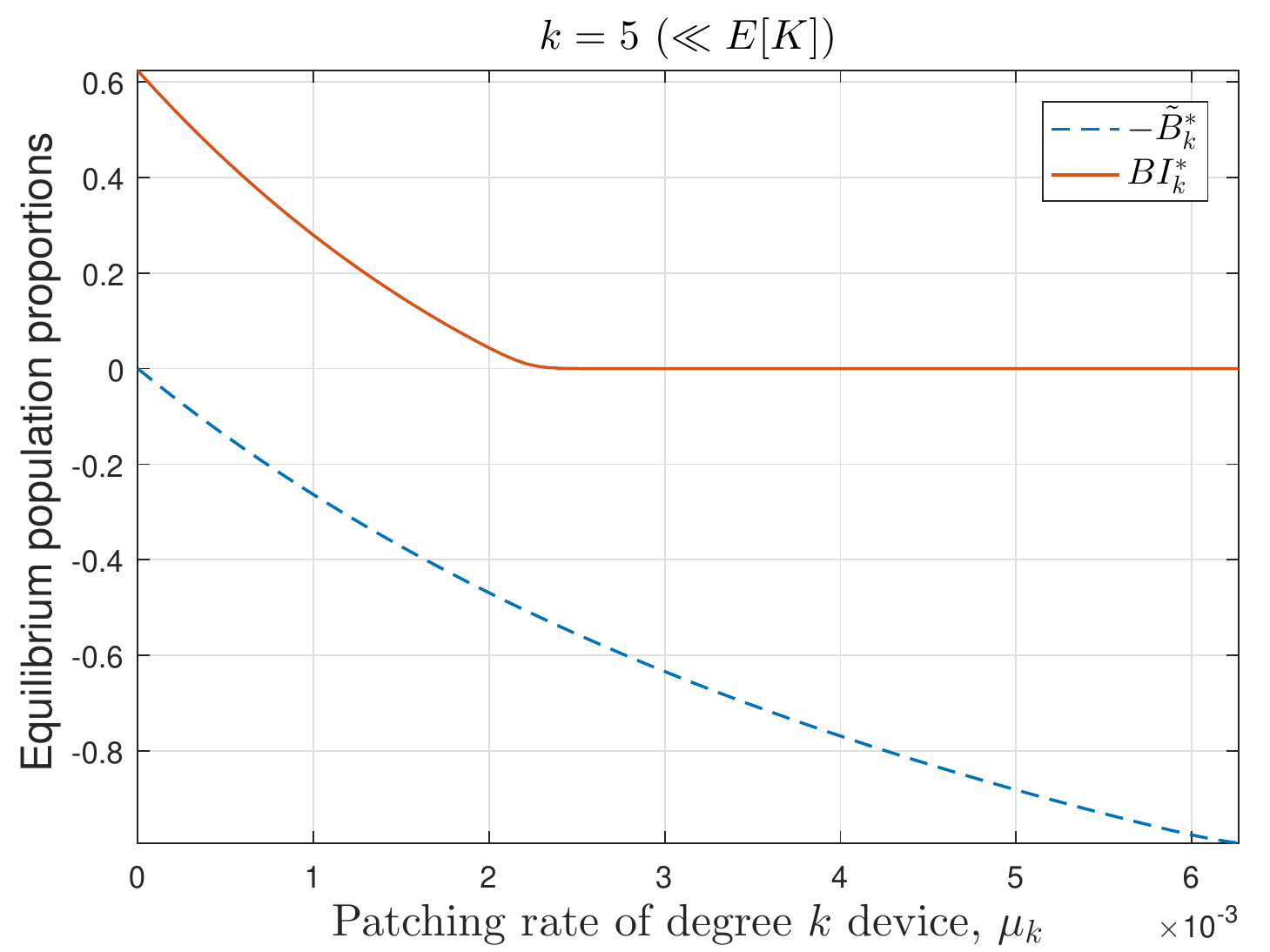} \label{Fig:degree_small}} \ \
\subfloat[Equilibrium populations against patching rate for large degree devices.]{\includegraphics[width = 3in]{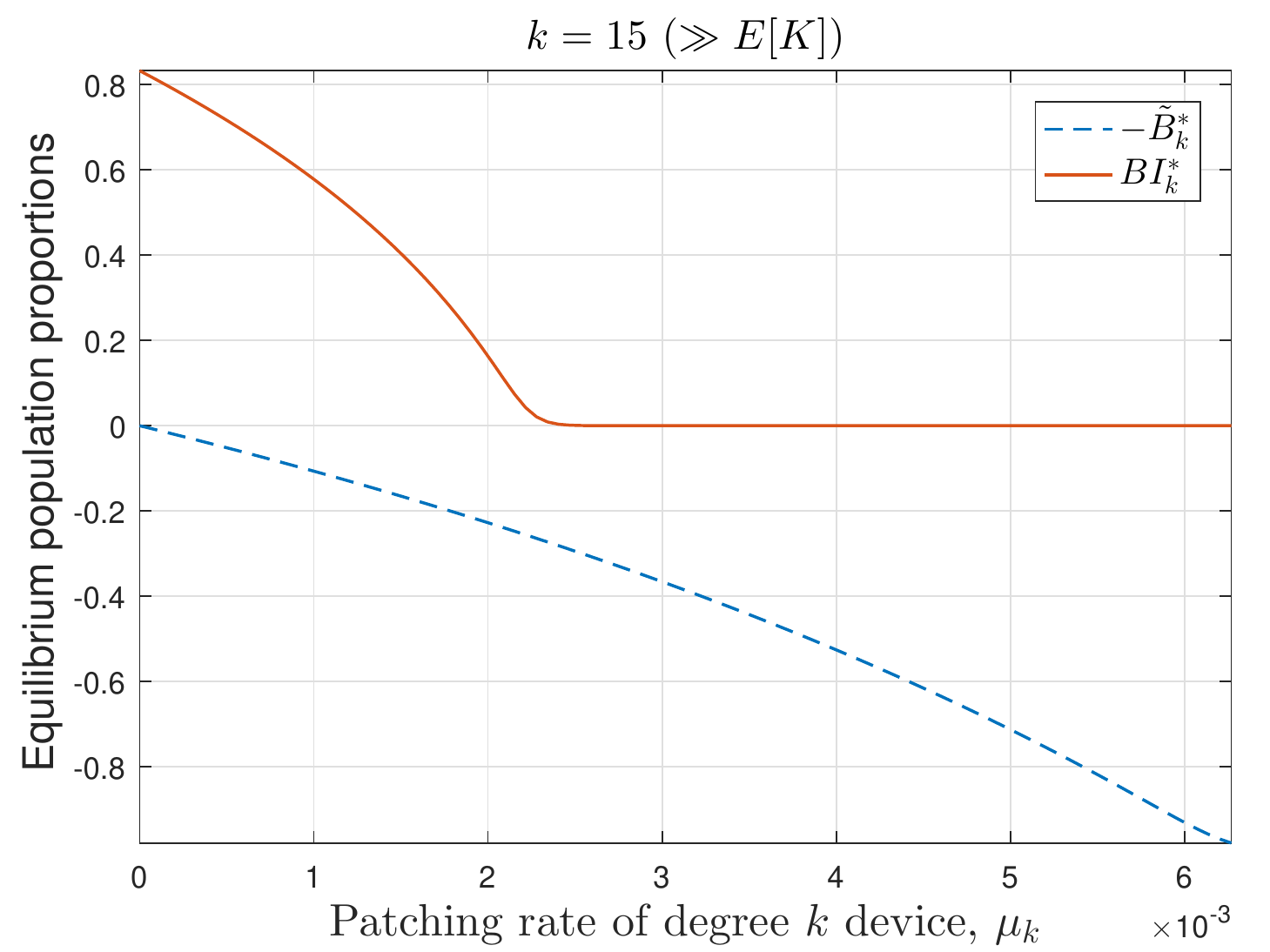} \label{Fig:degree_large}}
\caption{Curvature analysis of equilibrium population processes for different degree devices.}
\label{Fig:curvature}
\end{figure*}

We can observe that $\frac{d \tilde{B}_k^*}{d \mu_k} = \frac{k \sigma_1 - \mu_k k \sigma_1^{\prime}}{(\mu_k + k \sigma_1)^2}$ and $\frac{d^2 \tilde{B}_k^*}{d \mu_k^2} = \frac{(\mu_k + k \sigma_1)\left( (\mu_k + k \sigma_1)(- \mu_k k \sigma_1^{\prime \prime}) - 2(1 + k \sigma_1^{\prime})(k \sigma_1 - \mu_k k \sigma_1^{\prime})\right)}{(\mu_k + k \sigma_1)^3}$, where $\sigma_1^{\prime} = \frac{d \sigma_1(\mu_k)}{d \mu_k} = -\frac{1}{\mathbb{E}[K]}$ and $\sigma_1^{\prime \prime} = \frac{d^2 \sigma_1(\mu_k)}{d \mu_k^2} = 0$. The denominator of $\frac{d^2 \tilde{B}_k^*}{d \mu_k^2}$ is always positive and the numerator evaluates to $-2 (\mu_k + k \sigma_1)\left(1 - \frac{k}{\mathbb{E}[K]} \right) \left( k \sigma_1 + \frac{k \mu_k}{\mathbb{E}[K]}\right)$. Therefore, it is clear that $\frac{d^2 \tilde{B}_k^*}{d \mu_k^2} < 0$ if $k < \mathbb{E}[K]$ and vice versa. Therefore, we can conclude that $\tilde{B}_k$ evaluated at equilibrium is concave for $k < \mathbb{E}[K]$ and convex otherwise. Similarly, for $BI_k$, it can be shown that $\frac{d^2 BI_k^*}{d \mu_k^2}$ experiences a change in sign with $k$, which is hard to characterize analytically but the change point can be proved to be different than $\mathbb{E}[K]$. In order to demonstrate the change in curvature of the equilibrium populations, we plot the respective equilibrium populations of un-compromised devices and informed bots in Fig.~\ref{Fig:curvature} for different values of $k$. Note that with an increasing patching rate, the un-compromised device population increases until it reaches 1 ($- \tilde{B}_k$ is plotted in Fig.~\ref{Fig:curvature}, which is decreasing to $-1$). However, on the other hand, the equilibrium population of informed bot devices decreases until it reaches 0. Furthermore, the informed bot device population diminishes completely with a much smaller patching rate that is required to make the network completely un-compromised. These equilibrium populations have been plotted with mean device degree $\mathbb{E}[K] = 9.4$ and it can be observed that the curvature of the constraints is different if the degree is small, i.e., $k = 5$, than when it is large, i.e., $k =15$.

\section{Proof of Lemma~\ref{lemma_max_threshold}}\label{proof_max_threshold}
From Appendix~\ref{Patching_limit_proof}, it can be concluded that $\hat{\mu}_k = \frac{\rho \gamma_b \gamma_c p \mathbb{E}[K] - \rho \gamma_b \beta}{\gamma_c + \rho \gamma_b}$ can completely eradicate equilibrium population of informed bots of degree $k$. However, at this patching rate, the population of un-compromised devices can be obtained as $\tilde{B}_k^*(\hat{\mu}_k) = \frac{\mathbb{E}[K](p \gamma_c \mathbb{E}[K] - \beta)}{k(\beta + \mathbb{E}[K]p \rho \gamma_b) + \mathbb{E}[K](\mathbb{E}[K]p \gamma_c - \beta)}$. Since $\tilde{B}_k^*(\hat{\mu}_k)$ is a convex function of $k$, $\sum_{k=1}^{\infty} \tilde{B}_k^*(\hat{\mu}_k) = \mathbb{E}[\tilde{B}_k^*(\hat{\mu}_k)] \geq \tilde{B}_{\mathbb{E}[K]}^*(\hat{\mu}_k)$ (Using Jensen's inequality~\cite{jensen}). It results in $\mathbb{E}[\tilde{B}_k^*(\hat{\mu}_k)] \geq \frac{\mathbb{E}[K]p\gamma_c - \beta}{\mathbb{E}[K]p (\rho \gamma_b + \gamma_c)}$. Knowing that $\tilde{B}_k^*$ is an increasing function of $\mu_k$, we can deduce that if $\tau_{\tilde{B}} \geq \frac{\mathbb{E}[K]p\gamma_c - \beta}{\mathbb{E}[K]p (\rho \gamma_b + \gamma_c)}$, then it requires a patching rate higher than $\hat{\mu}_k$. This implies that $BI_k^*$ will be zero at the optimal patching rate. Hence, the constraint~\eqref{const_BI_original} will always be satisfied if $\tau_{\tilde{B}}$ is sufficiently high and therefore, we can effectively remove it from the optimization problem. This phenomenon can also be observed from Fig.~\ref{Fig:curvature} where the equilibrium population of informed bots diminishes to zero much earlier than the equilibrium proportion of un-compromised devices.

\section{} \label{duality_gap}
To prove that the duality gap for the optimization problem formulated in~\cref{obj_original,const_B_tilde_original,const_BI_original} is zero, we invoke a key result from~\cite{dual_methods}. An adaptation of its statement is provided as follows: Consider the primal optimization problem of the form $\text{minimize} \ \sum_{k=1}^{k_{\max}} f_k(x_k)$ subject to $\sum_{k=1}^{k_{\max}} h_k(x_k) \leq \mathbf{P}$, where $f_k(.)$ is a scalar function, $h_k(.)$ is a vector function, and $\mathbf{P}$ is a vector of constraints. Both $f_k(.)$ and $h_k(.)$ may not necessarily be convex. Now, let $\mathbf{x}$ and $\mathbf{y}$ be the optimal solutions to be problem with $\mathbf{P} = P_{\text{x}}$ and $\mathbf{P} = P_{\text{y}}$ respectively. Then, for $\nu \in [0,1]$, if there exists $\mathbf{z}$ such that $\sum_{k=1}^{k_{\max}} h_k(z_k) \leq \nu P_{\text{x}} + (1 - \nu)P_{\text{y}}$ and $\sum_{k=1}^{k_{\max}} f_k(z_k) \leq \nu \sum_{k=1}^{k_{\max}} f_k(x_k) + (1 - \nu)\sum_{k=1}^{k_{\max}} f_k(y_k)$, then the duality gap is zero leading to the same solution for the primal and dual problems. For more details, the readers are referred to~\cite{dual_methods} and references therein. Now, for the problem considered in this paper, the objective is strictly convex while the constraints may not necessarily be convex. Assuming that we are considering the feasible regime for $\boldsymbol{\mu}$ as defined in Corollary~\ref{Patching_limit} and $\boldsymbol{\mu}_{\text{x}}$, $\boldsymbol{\mu}_{\text{y}}$ are the optimal patching rates corresponding to threshold vectors $P_{\text{x}}$ and $P_{\text{y}}$. First, assume that only the constraint~\eqref{const_B_tilde_original} is active, i.e., $P_{\text{x}} = \tau_{\tilde{B}}^{\text{x}}$ and $P_{\text{y}} = \tau_{\tilde{B}}^{\text{y}}$ are scalars. Since $\tilde{B}_k^*(.)$ is strictly monotone, so if $\tau_{\tilde{B}}^{\text{x}}  > \tau_{\tilde{B}}^{\text{y}}$, then the optimal $\mu_{\text{x},k} > \mu_{\text{y},k}, \ \forall k$. Therefore, there exists an interior point $\boldsymbol{\mu}_{\text{z}} = \{\mu_{z,k} : \min(\mu_{\text{x},k}, \mu_{\text{y},k}) \leq z_k \leq \max(\mu_{\text{x},k}, \mu_{\text{y},k}), \ \forall k \}$ for which $- \sum_{k=1}^{k_{\max}} \tilde{B}_k^*(\mu_{z,k}) \leq - \nu \tau^{\text{x}}_{\tilde{B}} - (1 - \nu)\tau^{\text{y}}_{\tilde{B}}$. From the convexity of $\phi_k(.)$ in the objective in~\eqref{obj_original}, it is clear that $\sum_{k=1}^{k_{\max}} \phi_k(\mu_{\text{z},k}) \pi_k \leq \nu \sum_{k=1}^{k_{\max}} \phi_k(\mu_{\text{x},k}) \pi_k + (1-\nu) \sum_{k=1}^{k_{\max}} \phi_k(\mu_{\text{y},k}) \pi_k$. This implies that the duality gap of the problem is zero. Now, when both constraints~\eqref{const_BI_original} and~\eqref{const_B_tilde_original} are active, the argument still applies since both $- \tilde{B}^*_k(.)$ and $BI_k^*$ are strictly decreasing functions of the arguments which guarantees the existence of an interior point corresponding to every linear combination of $P_{\text{x}}$ and $P_{\text{y}}$ The convexity of the objective function subsequently completes the proof.

\bibliographystyle{IEEEtran}
\vspace{-0.0in}
\bibliography{references}

\begin{thebibliography}{10}
\providecommand{\url}[1]{#1}
\csname url@samestyle\endcsname
\providecommand{\newblock}{\relax}
\providecommand{\bibinfo}[2]{#2}
\providecommand{\BIBentrySTDinterwordspacing}{\spaceskip=0pt\relax}
\providecommand{\BIBentryALTinterwordstretchfactor}{4}
\providecommand{\BIBentryALTinterwordspacing}{\spaceskip=\fontdimen2\font plus
\BIBentryALTinterwordstretchfactor\fontdimen3\font minus
  \fontdimen4\font\relax}
\providecommand{\BIBforeignlanguage}[2]{{%
\expandafter\ifx\csname l@#1\endcsname\relax
\typeout{** WARNING: IEEEtran.bst: No hyphenation pattern has been}%
\typeout{** loaded for the language `#1'. Using the pattern for}%
\typeout{** the default language instead.}%
\else
\language=\csname l@#1\endcsname
\fi
#2}}
\providecommand{\BIBdecl}{\relax}
\BIBdecl

\bibitem{iot_ran}
S.~Al-Sarawi, M.~Anbar, K.~Alieyan, and M.~Alzubaidi, ``{Internet} of things
  ({IoT}) communication protocols: Review,'' in \emph{8th Intl. Conf. Inf.
  Technol. (ICIT 2017)}, May 2017, pp. 685--690.

\bibitem{echo}
\BIBentryALTinterwordspacing
{Amazon Echo}. [Online]. Available:
  \url{https://www.amazon.com/Amazon-Echo-And-Alexa-Devices/b?ie=UTF8&node=9818047011}
\BIBentrySTDinterwordspacing

\bibitem{g_home}
\BIBentryALTinterwordspacing
{Google Home}. [Online]. Available:
  \url{\https://store.google.com/us/product/google_home?hl=en-US}
\BIBentrySTDinterwordspacing

\bibitem{flaws}
A.~Tannenbaum, ``Why do {IoT} companies keep building devices with huge
  security flaws?'' Harvard Business Review, Apr. 2017.

\bibitem{cyber_security}
Y.~Dibrov, ``The {Internet} of things is going to change everything about
  cybersecurity,'' Harvard Business Review, Dec. 2017.

\bibitem{ddos_in_iot}
C.~Kolias, G.~Kambourakis, A.~Stavrou, and J.~Voas, ``{DDoS} in the {IoT}:
  Mirai and other botnets,'' \emph{Computer}, vol.~50, no.~7, pp. 80--84, 2017.

\bibitem{botnet}
M.~Feily, A.~Shahrestani, and S.~Ramadass, ``A survey of botnet and botnet
  detection,'' in \emph{3rd Intl. Conf. Emerging Security Inf. Sys. Technol.},
  June 2009, pp. 268--273.

\bibitem{botnet_survey_ppr}
G.~Vormayr, T.~Zseby, and J.~Fabini, ``Botnet communication patterns,''
  \emph{IEEE Commun. Surveys Tuts.}, vol.~19, no.~4, pp. 2768--2796, Fourth
  Quarter 2017.

\bibitem{ransom}
\BIBentryALTinterwordspacing
E.~Bertino and N.~Islam, ``Botnets and {Internet} of things security,''
  \emph{Computer}, vol.~50, no.~2, pp. 76--79, Feb. 2017. [Online]. Available:
  \url{doi.ieeecomputersociety.org/10.1109/MC.2017.62}
\BIBentrySTDinterwordspacing

\bibitem{mmirai}
\BIBentryALTinterwordspacing
M.~A. et~al, ``Understanding the {Mirai} botnet,'' in \emph{Proceedings of the
  26th USENIX Security Symposium}, 2017. [Online]. Available:
  \url{https://www.usenix.org/conference/usenixsecurity17/technical-sessions/presentation/antonakakis}
\BIBentrySTDinterwordspacing

\bibitem{variant}
P.~Moriuchi and S.~Chohan, ``Mirai-variant {IoT} botnet used to target
  financial sector in january 2018,'' Insikt Group, Apr. 2018.

\bibitem{wifi_botnets}
M.~Knysz, X.~Hu, Y.~Zeng, and K.~G. Shin, ``Open {WiFi} networks: Lethal
  weapons for botnets?'' in \emph{Proc. IEEE Intl. Conf. Comput. Commun.
  (INFOCOM 2012)}, Orlando, FL, USA, Mar. 2012, pp. 2631--2635.

\bibitem{drone_botmaster}
T.~Reed, J.~Geis, and S.~Dietrich, ``{SkyNET}: A 3{G}-enabled mobile attack
  drone and stealth botmaster,'' in \emph{Proc. 5th USENIX Conf. on Offensive
  Technologies}, ser. WOOT'11.\hskip 1em plus 0.5em minus 0.4em\relax Berkeley,
  CA, USA: USENIX Association, 2011.

\bibitem{wdos}
K.~Pelechrinis, M.~Iliofotou, and S.~V. Krishnamurthy, ``Denial of service
  attacks in wireless networks: The case of jammers,'' \emph{IEEE Commun.
  Surveys Tuts.}, vol.~13, no.~2, pp. 245--257, Second Quarter 2011.

\bibitem{mac_whitepaper}
``Can wireless {LAN} denial of service attacks be prevented? understanding
  {WLAN DoS} vulnerabilities \& practical countermeasures,'' Motorola Inc.,
  White Paper, 2009.

\bibitem{soa3}
N.~Vlajic and D.~Zhou, ``{IoT} as a land of opportunity for {DDoS} hackers,''
  \emph{Computer}, vol.~51, no.~7, pp. 26--34, Jul. 2018.

\bibitem{tifs_2_internet}
Q.~Wang, Z.~Chen, and C.~Chen, ``On the characteristics of the worm infection
  family tree,'' \emph{IEEE Trans. Inf. Forensics and Security}, vol.~7, no.~5,
  pp. 1614--1627, Oct 2012.

\bibitem{measurement}
J.~Kim, S.~Radhakrishnan, and S.~K. Dhall, ``Measurement and analysis of worm
  propagation on {Internet} network topology,'' in \emph{Proc.13th Intl. Conf.
  Computer Commun. Netw. (IEEE Cat. No.04EX969)}, Oct. 2004, pp. 495--500.

\bibitem{simulation}
K.~Channakeshava, D.~Chafekar, K.~Bisset, V.~S.~A. Kumar, and M.~Marathe,
  ``Epinet: A simulation framework to study the spread of malware in wireless
  networks,'' in \emph{Proc. 2nd Intl. Conf. Simulation Tools and Techniques},
  ser. Simutools '09.\hskip 1em plus 0.5em minus 0.4em\relax Brussels, Belgium,
  Belgium: Institute for Computer Sciences, Social-Informatics and
  Telecommunications Engineering (ICST), 2009.

\bibitem{soa4}
D.~Yin, L.~Zhang, and K.~Yang, ``A {DDoS} attack detection and mitigation with
  software-defined internet of things framework,'' \emph{IEEE Access}, vol.~6,
  pp. 24\,694--24\,705, 2018.

\bibitem{iot_epidemic}
\BIBentryALTinterwordspacing
J.~A. Jerkins and J.~Stupiansky, ``Mitigating {IoT} insecurity with inoculation
  epidemics,'' in \emph{Proceedings of the ACMSE 2018 Conference}, ser. ACMSE
  '18.\hskip 1em plus 0.5em minus 0.4em\relax New York, NY, USA: ACM, 2018, pp.
  4:1--4:6. [Online]. Available:
  \url{http://doi.acm.org/10.1145/3190645.3190678}
\BIBentrySTDinterwordspacing

\bibitem{tmc_botnets}
Z.~Lu, W.~Wang, and C.~Wang, ``On the evolution and impact of mobile botnets in
  wireless networks,'' \emph{IEEE Trans. Mobile Comput.}, vol.~15, no.~9, pp.
  2304--2316, Sep. 2016.

\bibitem{soa1}
J.~Xu, L.~Chen, K.~Liu, and C.~Shen, ``Designing security-aware incentives for
  computation offloading via device-to-device communication,'' \emph{IEEE
  Transactions on Wireless Communications}, vol.~17, no.~9, pp. 6053--6066,
  Sept. 2018.

\bibitem{soa2}
A.~A. Santos, M.~Nogueira, and J.~M.~F. Moura, ``A stochastic adaptive model to
  explore mobile botnet dynamics,'' \emph{IEEE Communications Letters},
  vol.~21, no.~4, pp. 753--756, Apr. 2017.

\bibitem{tifs_differential_game}
S.~Shen, H.~Li, R.~Han, A.~V. Vasilakos, Y.~Wang, and Q.~Cao, ``Differential
  game-based strategies for preventing malware propagation in wireless sensor
  networks,'' \emph{IEEE Trans Inf. Forensics and Security}, vol.~9, no.~11,
  pp. 1962--1973, Nov 2014.

\bibitem{twc_junaid}
M.~J. Farooq and Q.~Zhu, ``On the secure and reconfigurable multi-layer network
  design for critical information dissemination in the {Internet} of
  battlefield things {(IoBT)},'' \emph{IEEE Trans. Wireless Commun.}, vol.~17,
  no.~4, pp. 2618--2632, Apr. 2018.

\bibitem{wiopt_junaid}
------, ``Secure and reconfigurable network design for critical information
  dissemination in the internet of battlefield things {(IoBT)},'' in \emph{15th
  Intl. Symp. Model Optim. in Mobile, Ad Hoc, and Wireless Netw. (WiOpt 2017)},
  May 2017, pp. 1--8.

\bibitem{book_epidemiology}
F.~Brauer, P.~van~den Driessche, and E.~J.~Wu, \emph{Mathematical
  Epidemiology}.\hskip 1em plus 0.5em minus 0.4em\relax Springer, Berlin:
  Springer, 2008.

\bibitem{virus_spread}
\BIBentryALTinterwordspacing
A.~L. Lloyd and R.~M. May, ``How viruses spread among computers and people,''
  \emph{Science}, vol. 292, no. 5520, pp. 1316--1317, 2001. [Online].
  Available: \url{http://science.sciencemag.org/content/292/5520/1316}
\BIBentrySTDinterwordspacing

\bibitem{rumour_dynamics}
\BIBentryALTinterwordspacing
Y.~Moreno, M.~Nekovee, and A.~F. Pacheco, ``Dynamics of rumor spreading in
  complex networks,'' \emph{Phys. Rev. E}, vol.~69, p. 066130, Jun 2004.
  [Online]. Available:
  \url{https://link.aps.org/doi/10.1103/PhysRevE.69.066130}
\BIBentrySTDinterwordspacing

\bibitem{population_processes}
J.~F.~C. Kingman, ``Markov population processes,'' \emph{Journal of Applied
  Probability}, vol.~6, no.~1, pp. 1--18, 1969.

\bibitem{sg}
D.~Stoyan, W.~S. Kendall, and J.~Mecke, \emph{Stochastic geometry and its
  applications}, ser. Wiley series in probability and mathematical
  statisitics.\hskip 1em plus 0.5em minus 0.4em\relax Chichester, W. Sussex,
  New York: Wiley, 1987.

\bibitem{nyc_hotspot_data}
\BIBentryALTinterwordspacing
{NYC OpenData, NYC Wi-Fi Hotspot Locations}. [Online]. Available:
  \url{https://data.cityofnewyork.us/Social-Services/NYC-Wi-Fi-Hotspot-Locations/a9we-mtpn}
\BIBentrySTDinterwordspacing

\bibitem{aloha}
\BIBentryALTinterwordspacing
N.~Abramson, ``{THE ALOHA SYSTEM}: Another alternative for computer
  communications,'' in \emph{Proceedings of the November 17-19, 1970, Fall
  Joint Computer Conference}, ser. AFIPS '70 (Fall).\hskip 1em plus 0.5em minus
  0.4em\relax New York, NY, USA: ACM, 1970, pp. 281--285. [Online]. Available:
  \url{http://doi.acm.org/10.1145/1478462.1478502}
\BIBentrySTDinterwordspacing

\bibitem{aloha_SG}
S.~Weber, J.~G. Andrews, and N.~Jindal, ``An overview of the transmission
  capacity of wireless networks,'' \emph{IEEE Trans. Commun.}, vol.~58, no.~12,
  pp. 3593--3604, Dec. 2010.

\bibitem{outage_aloha}
M.~Haenggi, ``Outage, local throughput, and capacity of random wireless
  networks,'' \emph{IEEE Trans. Wireless Commun.}, vol.~8, no.~8, pp.
  4350--4359, Aug. 2009.

\bibitem{kaynia_outage}
M.~Kaynia and N.~Jindal, ``Performance of {ALOHA} and {CSMA} in spatially
  distributed wireless networks,'' in \emph{2008 IEEE Intl. Conf. Commun.}, May
  2008, pp. 1108--1112.

\bibitem{epidemics}
R.~Pastor-Satorras, C.~Castellano, P.~Van~Mieghem, and A.~Vespignani,
  ``Epidemic processes in complex networks,'' \emph{Rev. Mod. Phys.}, vol.~87,
  pp. 925--979, Aug. 2015.

\bibitem{nonuniform_transmission}
\BIBentryALTinterwordspacing
C.~yi~Xia, Z.~Wang, J.~Sanz, S.~Meloni, and Y.~Moreno, ``Effects of delayed
  recovery and nonuniform transmission on the spreading of diseases in complex
  networks,'' \emph{Physica A: Statistical Mechanics and its Applications},
  vol. 392, no.~7, pp. 1577 -- 1585, 2013. [Online]. Available:
  \url{http://www.sciencedirect.com/science/article/pii/S0378437112010084}
\BIBentrySTDinterwordspacing

\bibitem{dual_methods}
W.~Yu and R.~Lui, ``Dual methods for nonconvex spectrum optimization of
  multicarrier systems,'' \emph{IEEE Trans. Commun.}, vol.~54, no.~7, pp.
  1310--1322, July 2006.

\bibitem{num_chiang}
D.~P. Palomar and M.~Chiang, ``A tutorial on decomposition methods for network
  utility maximization,'' \emph{IEEE J. Sel. Areas Commun.}, vol.~24, no.~8,
  pp. 1439--1451, Aug. 2006.

\bibitem{dual_decomposition}
L.~Xiao, M.~Johansson, and S.~P. Boyd, ``Simultaneous routing and resource
  allocation via dual decomposition,'' \emph{IEEE Trans. Commun.}, vol.~52,
  no.~7, pp. 1136--1144, July 2004.

\bibitem{jensen}
Z.~Cvetkovski, \emph{Inequalities: Theorems, Techniques and Selected Problems},
  Springer, Berlin, Heidelberg, 2012, ch. Convexity, Jensen's Inequality, pp.
  69--77.

\end{thebibliography}

\begin{IEEEbiography}
	[{\includegraphics[width=1in,height=1.25in,clip,keepaspectratio]{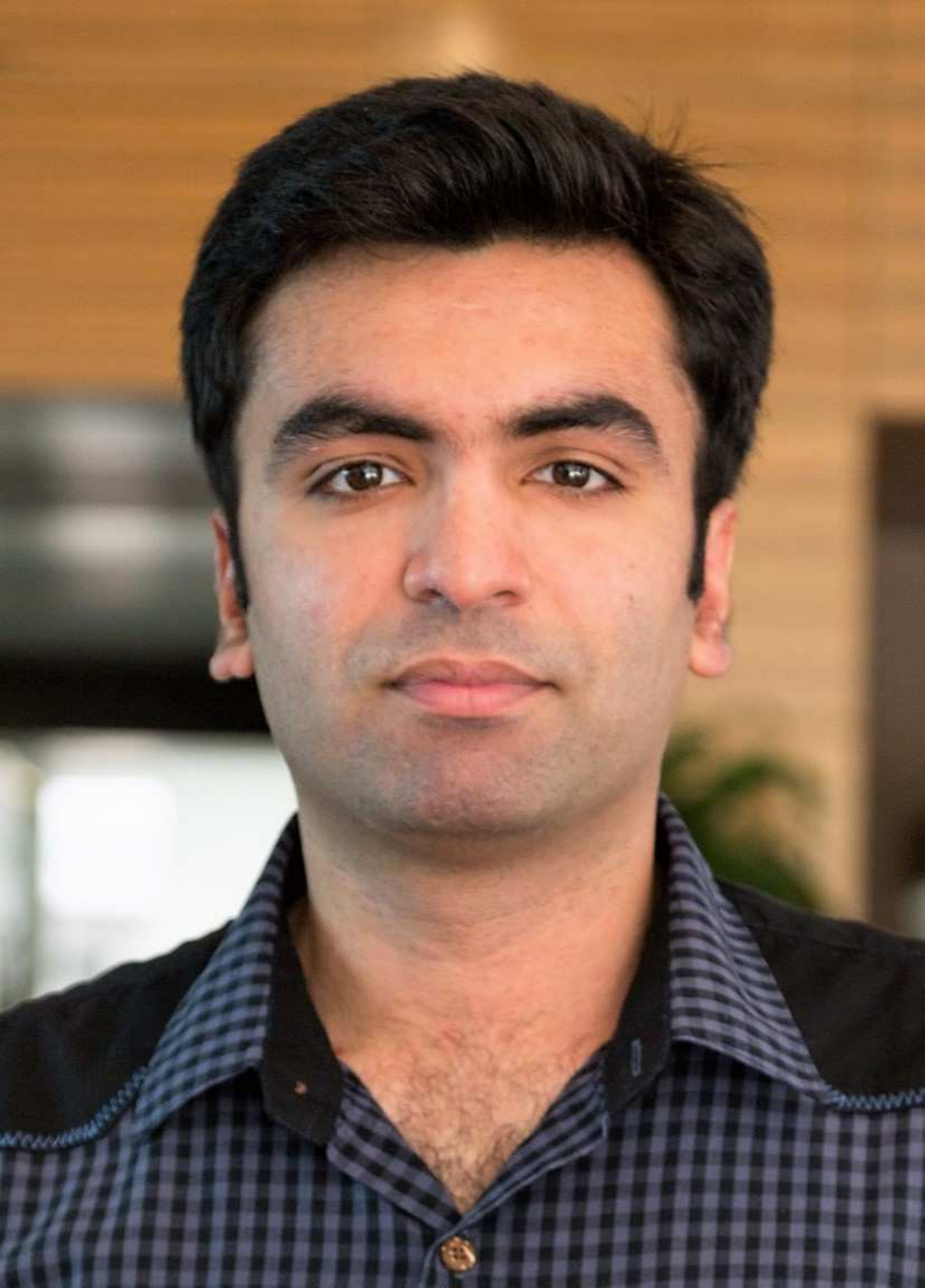}}]{Muhammed Junaid Farooq} received the B.S. degree in electrical engineering from the School of Electrical Engineering and Computer Science (SEECS), National University of Sciences and Technology (NUST), Islamabad, Pakistan, the M.S. degree in electrical engineering from the King Abdullah University of Science and Technology (KAUST), Thuwal, Saudi Arabia, in 2013 and 2015, respectively. Then, he was a Research Assistant with the Qatar Mobility Innovations Center (QMIC), Qatar Science and Technology Park (QSTP), Doha, Qatar. Currently, he is a PhD student at the Tandon School of Engineering, New York University (NYU), Brooklyn, New York. His research interests include modeling, analysis and optimization of wireless communication systems, cyber-physical systems, and the Internet of things. He is a recipient of the President's Gold Medal for academic excellence from NUST, the Ernst Weber Fellowship Award for graduate studies and the Athanasios Papoulis Award for graduate teaching excellence from the department of Electrical \& Computer Engineering (ECE) at NYU Tandon School of Engineering.
\end{IEEEbiography}

\begin{IEEEbiography}
	[{\includegraphics[width=1in,height=1.25in,clip,keepaspectratio]{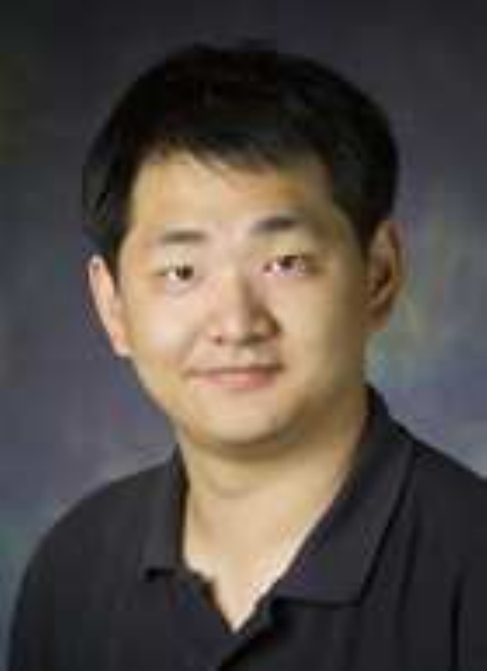}}]{Quanyan Zhu} (S'04, M'12) received B. Eng. in Honors Electrical Engineering from McGill University in 2006, M.A.Sc. from University of Toronto in 2008, and Ph.D. from the University of Illinois at Urbana-Champaign (UIUC) in 2013. After stints at Princeton University, he is currently an assistant professor at the Department of Electrical and Computer Engineering, New York University. He is a recipient of many awards including NSERC Canada Graduate Scholarship (CGS), Mavis Future Faculty Fellowships, and NSERC Postdoctoral Fellowship (PDF). He spearheaded and chaired INFOCOM Workshop on Communications and Control on Smart Energy Systems (CCSES), and Midwest Workshop on Control and Game Theory (WCGT). His current research interests include Internet of things, cyber-physical systems, security and privacy, and system and control.
\end{IEEEbiography}

\end{document}